\RequirePackage{fix-cm}
\documentclass[smallextended]{svjour3}       % onecolumn (second format)
\smartqed  % flush right qed marks, e.g. at end of proof
\usepackage{graphicx}
\journalname{Empirical Software Engineering}
\graphicspath{{figures/}}
\DeclareGraphicsExtensions{.pdf,.png}

\usepackage{natbib}
\setcitestyle{authoryear}

\usepackage{multirow}

\usepackage[usenames,dvipsnames,svgnames,table]{xcolor} % http://en.wikibooks.org/wiki/LaTeX/Colors

\usepackage[export]{adjustbox} % draw boxes around figures
\usepackage{amssymb,amsmath}

\usepackage[hyphens]{url}

\usepackage{enumitem}

\usepackage{calc}

\usepackage[framemethod=tikz]{mdframed}
\newmdenv[
  innerleftmargin=7pt,
  innerrightmargin=7pt,
  skipabove=\baselineskip,
  skipbelow=\baselineskip,
  tikzsetting={draw=black,dashed,line width=0.5pt,dash pattern = on 4pt off 2pt},
  linecolor=white,
  backgroundcolor=white
]{dashedbox}
\newmdenv[
  innerleftmargin=7pt,
  innerrightmargin=7pt,
  skipabove=\baselineskip,
  skipbelow=\baselineskip,
  tikzsetting={draw=black, line width=0.5pt},
  linecolor=black,
  backgroundcolor=white
]{normalbox}
\newmdenv[
  topline=false,
  bottomline=false,
  rightline=false,
  skipabove=\topsep,
  skipbelow=\topsep,
  innertopmargin=0pt,
  innerbottommargin=0pt,
  innerleftmargin=7pt,
  innerrightmargin=0pt,
  tikzsetting={draw=black, line width=3pt},
  linecolor=black,
  backgroundcolor=white
]{verticalline}

\usepackage{listings}

\definecolor{light-gray}{gray}{0.97}
\definecolor{gray}{rgb}{0.4,0.4,0.4}
\definecolor{darkblue}{rgb}{0.0,0.0,0.6}
\definecolor{cyan}{rgb}{0.0,0.6,0.6}

\lstnewenvironment{java} {\lstset{
  language=Java
}}{}

\lstnewenvironment{code} {\lstset{
}}{}

\lstnewenvironment{regex} {\lstset{
  alsoletter=\\()\{\}.,
  morekeywords={String, long, if, return, format, B, int, Math, log}
}}{}

\usepackage{hyperref}

\definecolor{VeryLightGray}{rgb}{0.92,0.92,0.92}
\newcommand{\graybg}{\cellcolor{VeryLightGray}}

\usepackage{nicefrac}

\begin{document}

\title{Usage and Attribution of Stack Overflow\\ Code Snippets in GitHub Projects}
%\subtitle{}

%\titlerunning{Short form of title}        % if too long for running head

\author{Sebastian Baltes \and Stephan Diehl}

%\authorrunning{Short form of author list} % if too long for running head

\institute{	Sebastian Baltes
		\at University of Trier, Germany \\
		Orcid: 0000-0002-2442-7522\\
		\email{research@sbaltes.com}
           \and
           	Stephan Diehl
		 \at University of Trier, Germany \\
		 Orcid: 0000-0002-4287-7447\\
           	\email{diehl@uni-trier.de}  
}

\date{Received: 19 Januar 2018 / Accepted: 14 August 2018}
% The correct dates will be entered by the editor

\maketitle

\begin{abstract}
Stack Overflow (SO) is the most popular question-and-answer website for software developers, providing a large amount of copyable code snippets.
Using those snippets raises maintenance and legal issues. SO's license (CC BY-SA 3.0) requires attribution, i.e., referencing the original question or answer, and requires derived work to adopt a compatible license. While there is a heated debate on SO's license model for code snippets and the required attribution, little is known about the extent to which snippets are copied from SO without proper attribution. 
We present results of a large-scale empirical study analyzing the usage and attribution of non-trivial Java code snippets from SO answers in public GitHub (GH) projects.
We followed three different approaches to triangulate an estimate for the ratio of unattributed usages and conducted two online surveys with software developers to complement our results.
 For the different sets of projects that we analyzed, the ratio of projects containing files with a reference to SO varied between 3.3\% and 11.9\%.
We found that at most 1.8\% of all analyzed repositories containing code from SO used the code in a way compatible with CC BY-SA 3.0.
Moreover, we estimate that at most a quarter of the copied code snippets from SO are attributed as required. 
Of the surveyed developers, almost one half admitted copying code from SO without attribution and about two thirds were not aware of the license of SO code snippets and its implications.
% empirical study \and
\keywords{code snippets \and licensing \and stack overflow \and github \and online survey \and mining software repositories}
% \PACS{PACS code1 \and PACS code2 \and more}
% \subclass{MSC code1 \and MSC code2 \and more}
\end{abstract}

\section{Introduction}
\label{sec:introduction}

% Official SO data dump, released 2017-12-01
% select count(distinct Id) from Posts where PostTypeId in (1, 2);
% 38,394,917 (questions + answers)
% select count(distinct Id) from Posts where PostTypeId in (1, 2) and Id in (select distinct PostId from PostHistory);
% 38,394,896 (questions + answers with history available)
% select count(distinct Id) from PostHistory where PostId in (select distinct Id from Posts where PostTypeId in (1, 2));
% 99,674,128 (versions of questions + answers including metadata edits)
% select count(distinct Id) from Posts where PostTypeId=1;
% 14,995,834 (questions)
% select count(distinct ParentId) from Posts where PostTypeId=2;
% 13,025,900 (questions with answer)
% select count(distinct Id) from Posts where PostTypeId=1 and AcceptedAnswerId is not null;
% 8,034,235 (questions with accepted answer)
% select count(distinct Id) from Posts where PostTypeId=2;
% 23,399,083 (answers)
% select count(distinct Id) from Users;
% 8,123,937 (users)

Stack Overflow (SO) is the most popular question-and-answer website for software developers.
As of December 2017, its public data dump~\citep{StackExchangeInc2017b} lists over 13 million answered questions and over 8 million registered users.
Many answers contain code snippets together with explanations~\citep{YangHussainOthers2016}.
The availability of this large amount of code snippets lead to changes in software developers' behavior:
Nowadays, they regularly face the ``build or borrow'' question~\citep{BrandtDontchevaOthers2010}: Should they try to understand and solve an issue on their own or just copy and adapt a solution from SO?
Assuming that developers also copy and paste snippets from SO without trying to thoroughly understand them, \textit{maintenance issues} arise~\citep{ScalabrinoBavota2017}.
For instance, it may later be more difficult for developers to refactor or debug code that they did not write themselves. 
Moreover, if no link to the corresponding question or answer is added to the copied code, it is not possible to check the SO thread for a corrected or improved solution in case problems occur.
%\todo{developers with increasing frequency decide for the latter, the most copyable answers may get upvoted on SO, even if they are hard to understand or contain (minor) flaws.}

Beside possible maintainability implications, copying and pasting code from SO may also lead to \textit{licensing issues}: 				
All content on SO is currently licensed under the Creative Commons Attribution-ShareAlike 3.0 Unported license (CC BY-SA 3.0)~\citep{CreativeCommonsCorporation2007}, which allows to share and adapt the published content, but requires \textit{attribution} and demands contributions based on the content to be published under a compatible license (\textit{share-alike}).
Regarding the \textit{attribution} requirement, SO terms of service~\citep{StackExchangeInc2018b} stated---until May 2018---which information is required when content from SO is republished.
In particular, they required a link to the original post together with the names of the authors on SO (see Section~\ref{sec:attribution-requirements}). 

The \textit{share-alike} requirement of CC BY-SA 3.0 requires derived work to use a compatible license. It further requires adaptations of licensed content to add a credit identifying how the content is used. 
The license defines an adaptation as ``a work based upon'' the licensed content~\citep{CreativeCommonsCorporation2007}, which ``manifests sufficient new creativity to be copyrightable''~\citep{CreativeCommonsCorporation2017b}.
Regarding the licensing of such adaptations, CC BY-SA 3.0 restricts the way authors may distribute them, where distribution is defined as making the original work or an adaptation ``available to the public''.
It is only allowed to publish adaptations under the following licenses:
\begin{enumerate}
	\item CC BY-SA 3.0,
	\item a later version of CC BY-SA 3.0 (i.e., CC BY-SA 4.0),
	\item a ported version of CC BY-SA 3.0 (e.g., CC BY-SA 3.0 DE),
	\item a Creative Commons compatible license.
\end{enumerate}

% CC FAQ: "Recipients of the adaptation must comply with both the CC license on the original and your adapter’s license."
However, Creative Commons (CC) licenses are typically not used for software~\citep{Vendome2015} and there is currently no non-CC license that is considered share-alike compatible to CC BY-SA 3.0~\citep{CreativeCommonsCorporation2017}.
CC even recommends not to use CC licenses for software~\citep{CreativeCommonsCorporation2017b}, because, ``unlike software-specific licenses, CC licenses do not contain specific terms about the distribution of source code, which is often important to ensuring the free reuse and modifiability of software''.
They further state that ``it would be difficult to integrate CC-licensed work with other free software''.

% CC FAQ: You may offer the licensed material under other licenses in addition to the CC license (a practice commonly referred to as "dual licensing").
The situation is even more complicated, because code on Stack Overflow may have been copied from a source that has either a more permissive or a more restrictive license than SO (\textit{dual licensing}, see Section~\ref{sec:limitations}).
If such an external source does not provide a license at all, the author of the code still has the exclusive copyright and CC BY-SA 3.0 is the only license that applies for the code~\citep{GitHubInc2017, SojerHenkel2011}.
This situation makes the usage of code snippets from Stack Overflow problematic in terms of possible licensing conflicts (see Sections~\ref{sec:legal-situation} and \ref{sec:licensing-conflicts}).
%\todo{There are also contribution requirements in SO ToS: "Subscriber warrants, represents and agrees Subscriber has the right to grant Stack Exchange and the Network the rights set forth above. Subscriber represents, warrants and agrees that it will not contribute any Subscriber Content that..."}

%\todo{https://meta.stackexchange.com/questions/309797/were-examining-the-implementation-of-arbitration-in-the-2018-tos-update/309804\#309804}
%\todo{https://meta.stackexchange.com/questions/310061/electronic-opt-out-correcting-miscommunication-and-additional-questions-answer\#310061}
%\todo{https://meta.stackexchange.com/questions/310321/brace-yourselves-the-gdpr-is-coming}
%\todo{https://meta.stackexchange.com/questions/309786/how-can-we-opt-out-from-the-arbitration-clause-of-the-new-terms-of-service}
%\todo{https://meta.stackexchange.com/questions/273277/is-a-license-change-necessary}
%\todo{https://opensource.stackexchange.com/questions/6542/creative-commons-license-cc-by-sa-doesnt-work-for-commercial-content}
In May 2018, SO changed their terms of service, among other reasons, in response to the new \emph{European Union General Data Protection Regulation} (GDPR)~\citep{TimPost2018}.
With that change, the attribution requirements mentioned above were silently removed from the terms of service~\citep{StackExchangeInc2018c}.
However, the requirements are still (as of July 20, 2018) mentioned and linked in the footer of the website, which is visible for each thread, and in the help page.\footnote{\url{https://stackoverflow.com/help/licensing}}
Moreover, the terms of service now refer to version 4.0 of the CC BY-SA license, but the data dump is still licensed under version 3.0 (see Section~\ref{sec:licensing-conflicts} for information about the compatibility of versions 3.0 and 4.0).

%%%%%%%%%%%%%%%%%%%%%%%%%%%
% GH-Torrent on BigQuery (ghtorrent-bq:ght_2017_09_01)
% select count(*) from [ghtorrent-bq:ght_2017_09_01.projects] where deleted = false;
% 58,247,823 (all repos)
% select count(*) from [ghtorrent-bq:ght_2017_09_01.projects] where deleted = false and forked_from is null;
% 36,604,344 (non-fork repos)
GitHub (GH) is one of the most popular code hosting platforms with more than 58 million repositories (as of September 2017)~\citep{Gousios2017}.
It is not only used by developers for their personal projects, but also by large companies such as Google, Microsoft, or Facebook.
Since the source code of public GH projects is available online, copying and pasting code from SO posts into source code available on GH can be considered republication---the projects containing non-trivial code from Stack Overflow may even be considered adaptations of the copied code (see Section~\ref{sec:legal-situation} for details on when code is copyrightable).
Thus, the \textit{attribution} and the \textit{share-alike} requirements defined by CC BY-SA 3.0 apply.
If developers copy non-trivial code snippets from SO into their GH projects and fail to comply with those requirements, the license is terminated, which means that using the code may constitute \textit{copyright infringement}~\citep{St.Laurent2004, CreativeCommonsCorporation2017b}. % \citep{Carnes17}
%To adhere to SO's license, developers must attribute the origin when copying code snippets from SO.
%Moreover, regardless of the \textit{attribution} requirement, the \textit{share-alike} requirement implies that software build using non-trivial code snippets from SO must use a compatible license.  
% CC FAQ: "Similarly, if you are only distributing the material or adaptations of it within your company or organization, you do not have to comply with the attribution requirement." 
For closed source software projects, the \textit{attribution} requirement does not apply~\citep{CreativeCommonsCorporation2017b}.
However, the \textit{share-alike} requirement prevents using code from SO in closed source projects, it would only be allowed if the copied code is additionally licensed under a more permissive license.

To the best of our knowledge, there is currently no empirical evidence on how common it is to copy and paste non-trivial code snippets from SO into public GH projects without the required attribution ($\rightarrow$ RQ1).
It is also unclear how many of the projects using code from SO have a license conflict with Stack Overflow's license ($\rightarrow$ RQ2).
In the following, we present the research design and results of a large-scale analysis of the usage and attribution of Java code snippets from SO in public software projects hosted on GH.
We both analyze attributed usages and utilize three different approaches to estimate the ratio of unattributed usages.
To complement our results, we investigated if developers adhere to SO's attribution requirements ($\rightarrow$ RQ3) and conducted two surveys with software developers on their attribution practice and their awareness regarding the licensing of code from SO posts ($\rightarrow$ RQ4).

\section{Research Design}

The main goal of our research was to quantify the ratio of unattributed usages of code snippets from Stack Overflow in GitHub projects.
By usage we mean copying (and possibly slightly adapting) a code snippet from a post on SO and pasting it into a public GH project.
The following four research questions guided our research design:

\begin{itemize}[labelindent=\parindent, labelwidth=\widthof{\textbf{RQ1:}}, label=\textbf{RQ1:}, leftmargin=*, align=parleft, parsep=0pt, partopsep=0pt, topsep=1ex, itemsep=1ex]
\item[\textbf{RQ1:}] How often is code from Stack Overflow posts used in public GitHub projects without the required attribution?
\item[\textbf{RQ2:}] How often does the license of repositories containing code copied from Stack Overflow conflict with Stack Overflow's license?
\item[\textbf{RQ3:}] Do developers adhere to the attribution requirements defined in the Stack Overflow terms of service?
\item[\textbf{RQ4:}] Are software developers aware of the licensing of Stack Overflow code snippets and its implications?
\end{itemize}

We started our research with a preliminary survey to get first insights into developers's work practices regarding code snippets from SO (see Section~\ref{sec:preliminary-study}).
%To this end, we included questions on how developers use SO and GH, and in particular how they attribute code snippets, in a survey we conducted for another research project in the year 2015 (see Section~\ref{xxx}).
Our main research was then divided into three phases that focused on different files on GH, different code snippets from SO, and used different methods to triangulate an estimate for the ratio of unattributed usages (RQ1, see Sections~\ref{sec:phase1}, \ref{sec:phase2}, and \ref{sec:phase3}).
For all three phases, we retrieved the licenses of the repositories containing code from SO to assess their compatibility with CC BY-SA 3.0 (RQ2, see Section~\ref{sec:licensing-conflicts}).
To analyze the adherence to the SO attribution requirements, we manually analyzed a sample of Java files containing a link to an answer on SO (RQ3, see Section~\ref{sec:attribution-requirements}).
To assess the awareness of developers regarding the licensing of code from SO, we conducted a second online survey with GH project owners (RQ4, see Section~\ref{sec:awareness-survey}).

We used three main data sources to answer our research questions: The \textit{BigQuery GitHub data set}~\citep{GoogleCloudPlatform2017b},
the \textit{BigQuery GHTorrent data set}~\citep{Gousios2013, Gousios2017},
and the \textit{BigQuery Stack Overflow data set}~\citep{GoogleCloudPlatform2017}.
Google BigQuery provides a web-based console that allows to execute SQL queries on various public data sets, including the three data sets listed above.
For some aspects of our research, we retrieved additional information from the \textit{Stack Overflow data dump} released March 14, 2017~\citep{StackExchangeInc2017},
the \textit{GHTorrent data dump} released February 16, 2016~\citep{Gousios2013}, the \textit{GitHub API}~\citep{GitHubInc2017b},
and the \textit{Stack Exchange API}~\citep{StackExchangeInc2016b}.

We decided to restrict our analyses to Java, which is one of the most popular programming languages today~\citep{TIOBEsoftwareBV2017}.
Using the \textit{BigQuery SO data set}, we retrieved the frequency of question tags. The most common tag (as of March 2017) was \texttt{javascript} (1,339,747 questions), followed by \texttt{java} (1,223,171 questions).
Moreover, we used the \textit{BigQuery GHTorrent data set} to get the most common languages of non-fork projects.
Again, JavaScript was the most common language (2,194,750 projects) followed by Java (1,788,748).
According to GH's yearly report, Java was, considering the number of opened pull requests in 2017, the third most popular language on GH in that year (after JavaScript and Python)~\citep{GitHubInc2018}.

We chose Java over JavaScript, because Java has a unique file extension and is usually not embedded in other files (like JavaScript in HTML), which makes isolating Java code in SO posts and searching Java files on GH easier. 

In our research, we distinguish between \textit{attributed} and \textit{unattributed} usages of SO code snippets.
Attributed usages are relatively easy to detect due to the presence of a link to the content on SO.
To detect unattributed usages, we followed three different approaches:
In the first phase (see Section~\ref{sec:phase1}), we employed regular expressions to find copies of the snippets from the ten most frequently referenced Java answers on SO in all Java files in the \textit{BigQuery GH data set} (10 SO Java snippets, all Java files on GH).
In the second phase (see Section~\ref{sec:phase2}), we employed a code-clone detector to find clones of a sample of popular SO snippets in a sample of popular GH projects (227 SO Java snippets, 2,313 GH Java projects).
In the third phase (see Section~\ref{sec:phase3}), we searched for exact matches of as many SO snippets in as many GH Java files as computationally feasible with BigQuery (29,370 SO Java snippets, 1,720,587 GH Java files).
Our research mainly focused on finding type-1 clones of snippets, i.e., copied code that only varies in whitespace, layout, or comments~\citep{RoyCordyOthers2009}.
For such clones, we can be relatively sure that they have actually been copied from SO, assuming that the matches are not too short, the snippets are not too trivial, and there exists no other source.

In the following section, we briefly describe the legal situation, before we present the methods and results for each step of our research. We use framed boxes to summarize the results of each section and provide the raw data and all analysis scripts as supplementary material~\citep{Baltes2018e}.

\section{Legal Situation}
\label{sec:legal-situation}

In the following, we first describe the copyright status of SO code snippets, then classify SO's license as a strong copyleft license, and finally point to related discussions on different sites of the Stack Exchange network and related lawsuits.

\subsection{Copyright Status of Stack Overflow Code Snippets}

First of all, not all code snippets on SO are copyrightable.
Generally, ``copyright exists automatically whenever someone creates a work of authorship'' that is ``the author's intellectual creation''~\citep{Engelfriet2016}.
While this definition applies for software in general, many SO code snippets are only used to explain or demonstrate a solution, for example showing how to call a particular API.
In that scenario, the code would not be creative enough to be copyrightable~\citep{Engelfriet2016}.
During the famous \textit{Oracle v. Google} lawsuit (ongoing since 2012), Judge William H. Alsup ruled that APIs itself are generally not copyrightable~\citep{Alsup2012}.
However, this decision has been overturned by the Federal Circuit and the lawsuit is still ongoing~\citep{ElectronicFrontierFoundation2018}.

Arnoud Engelfriet, a Dutch IT law specialist, provides a rule of thumb that states ``if two programmers would provide substantially the same piece of code, the code is not creative under copyright law.''
He also mentions the often-quoted rule that ``anything less than ten lines of code is `trivial' and therefore not copyrighted'', but states that it is not grounded in any copyright legislation he is aware of.
Engelfriet concludes that ``a [Stack Overflow code] snippet that is more than one or two lines of standard function calls would typically be creative enough for copyright.'' and also argues against a fair use or quotation argument for such code snippets, mentioned for example by Jeff Atwood, the co-founder of SO~\citep{StackExchangeMeta2009}.

Since there exists no ``international standard for originality''~\citep{CreativeCommonsCorporation2017b} that defines when a code snippet is protected by copyright, we used popularity (phase 1), our own judgment (phase 2), and the snippets' length (phase 3) as proxy variables for their originality.
In a related study, we found that, as of December 2017, the mean size of code blocks on SO was 12 lines or 455 characters~\citep{BaltesDumaniOthers2018}, which supports our assumption that many snippets on SO are, at least according to their length, not trivial.

As outlined in the introduction, the code on SO may have been copied from a different source, with additional licensing and copyright implications.
We considered this in our research design by analyzing the external availability of the snippets (see Sections~\ref{sec:phase1}~and~\ref{sec:phase2}) and by excluding snippets that are also available from other sources (see Section~\ref{sec:phase3}).

\subsection{Classification of Stack Overflow's License}

Generally, one can distinguish between \textit{permissive} and \textit{copyleft} licenses.
Permissive licenses permit using the licensed source code in proprietary software without publishing changes or the derived work.
Examples for permissive licenses include the MIT, Apache, and BSD license families.
In contrast to that, copyleft licenses have a \textit{share-alike} requirement that requires either modifications to the licensed content or the complete derived work to be published under the same or a compatible license.
Examples for the former, weaker, copyleft licenses include the Mozilla and the Eclipse Public Licenses (e.g., MPL 2.0 and EPL 2.0); examples for the latter, stronger, copyleft licenses, which are sometime also called ``viral'' licenses~\citep{St.Laurent2004}, include the GNU General Public Licenses (e.g., GPL 2.0 and 3.0) and the Creative Commons Share-Alike Licenses (e.g., CC BY 2.0).
The licenses that apply for the content on SO (CC BY-SA 3.0 and 4.0) fall into the latter category and can thus be classified as strong copyleft licenses.

\subsection{Stack Overflow's License Change Attempt}

Licensing issues of source code posted on SO have been controversially discussed on different sites of the Stack Exchange network~\citep{StackExchangeMeta2009, StackExchangeMeta2013, StackExchangeMeta2015}.
% CC FAQ: In order for an adaptation to be protected by copyright, most national laws require the creator of the adaptation to add original expression to the pre-existing work. However, there is no international standard for originality, and the definition differs depending on the jurisdiction. Civil law jurisdictions (such as Germany and France) tend to require that the work contain an imprint of the adapter's personality. Common law jurisdictions (such as the U.S. or Canada), on the other hand, tend to have a lower threshold for originality, requiring only a minimal level of creativity and “independent conception.” Some countries approach originality completely differently. For example, Brazil's copyright code protects all works of the mind that do not fall within the list of works that are expressly defined in the statue as “unprotected works.” 
In December 2015, SO tried to switch to the more permissive MIT license for code snippets in new posts.
First, they planned to require attribution only upon request of the copyright holder or upon request of SO~\citep{StackExchangeMeta2015}
but after criticism from the community, they changed their proposal such that attribution would always be required~\citep{StackExchangeMeta2016}.
%In their description of the new license, SO stated that they consider an ``URL as a comment in [the] code'' to be reasonable attribution.
% an answer in another discussion~\citep{Huperniketes10}, where the author states that ``you cannot copy `snippets' of code via Fair Use'' but ``you can rewrite the way ideas [...]
%, systems (including algorithms), or factual information are expressed in those snippets''.
In January 2016, after a heated discussion, SO delayed the implementation of a new license and since then, no new proposal has been made.
Thus, as of July 2018, all source code posted on SO is licensed under CC BY-SA 3.0 (and 4.0) and the \textit{attribution} and \textit{share-alike} requirements apply.

\subsection{Related Lawsuits}

In the past, courts in the US and Europe ruled that open source licenses are enforceable contracts and that violations of open source licenses can be handled like copyright claims.
In the \textit{Jacobsen v. Katzer} lawsuit (2006--2010), the United States Court of Appeals for the Federal Circuit ruled that the terms and conditions of the Artistic License 1.0, including attribution, are ``enforceable copyright conditions''~\citep{White2008}.
In the \textit{Artifex v. Hancom} lawsuit (since 2016), the United States District Court for the Northern District of California denied a motion to dismiss~\citep{Corley2017}, arguing that a copyleft license like the GNU GPL can be treated like a legal contract.
This means that developers are able to sue when the terms of such a license are violated, e.g., when derived work is not shared under a compatible license (see the \textit{share-alike} requirement of CC BY-SA 3.0).
Moreover, it is possible to interdict the distribution of such derived work or claim monetary damages:
In 2004, the German District Court of Munich affirmed an injunctive relief interdicting the distribution of a software based on source code licensed under the GNU GPL, without complying with its license terms~\citep{KaessMullerOthers2004}.
In the United States, open source projects failing to comply with open source licenses can be targeted by DMCA takedown notices, which may force platforms like GH to remove projects that allegedly infringed copyright~\citep{Poteat2016}.
Recently, the Regional Court in Bochum, Germany, affirmed an obligation to pay compensation for damages in a case where source code licensed under the GNU GPL was used in violation of the license terms~\citep{AchteZivilkammer2016}.

Licensing issues may also be a risk in mergers and acquisitions of companies using source code licensed under a copyleft license~\citep{Cavaretta2015}.
A famous case was \textit{Free Software Foundation v. Cisco Systems}~\citep{SoftwareFreedomLawCenter2008}: Cisco acquired the networking company Linksys, which used GPL-licensed code in some of their products without publishing the source code.
After the Free Software Foundation (FSF) sued Cisco, they reached a settlement agreement, in which Cisco agreed to publish the source code and made an undisclosed financial contribution to the FSF~\citep{Wikipedia2017}.

\section{Preliminary Study}
\label{sec:preliminary-study}

We started our research with a preliminary study to get first insights into developers' practices regarding the usage and attribution of code snippets from Stack Overflow.

\subsection{Method:}

The preliminary study was part of an online survey we conducted in October 2015. For this survey, we contacted users who were active on both SO and GH.
To match users on both platforms, we followed the approach of Vasilescu et al., utilizing the MD5 hash value of users' email addresses~\citep{VasilescuFilkovOthers2013}.
We derived our sampling frame from the data dumps provided by \textit{Stack Exchange} (August 18, 2015)~\citep{StackExchangeInc2015} and \textit{GHTorrent} (September, 25 2015)~\citep{Gousios2013}.
To identify active users, we checked if they contributed to a question (asked, answered, or commented) on SO and committed to a project on GH since January 1, 2014.
This resulted in a sampling frame of 71,400 users from which we drew a random sample of 1,000 users.
Of the 1,000 contacted users, 122 responded (12.2\% response rate).
%\todo{add reference to Worse than Spam?}

\subsection{Results:}

Of all 122 respondents, 115 identified themselves as male, one as female and six did not provide their gender.
The majority of respondents (67\%) reported their main software development role to be \textit{software developer}, the second-largest group were \textit{software architects} (14\%).
The average age of participants was 28.9 years ($SD{=}9.1$) and they had an average programming experience of 11.8 years ($SD{=}6.7$).
Most participants answered from Europe (49\%) and North America (38\%).

We asked participants for what purpose they use SO and GH.
Most users answered that they use SO (98\%) and GH (66\%) for both private and work-related projects.
Almost one third of the respondents reported to use GH only for private projects (28\%).

A central question of the survey was: ``When was the last time you copied or adapted a code snippet from Stack Overflow?''
Most participants copied or adapted a snippet not more than one month ago (79\%) and over a third (39\%) not more than one week ago.
To get first insights into the attribution practice, we asked how they referred to the corresponding SO question or answer when they copied or adapted the snippet.
Half of the respondents (49\%) ``just copied/adapted the code snippet without any reference'', the others ``added a source code comment with a link to the Stack Overflow question/answer'' (40\%) or referred to SO in another way, e.g., in a commit message (9\%). Two participants did not answer this question.
%We provide the questionnaire and the closed-ended responses as supplementary material~\citep{Baltes2018e}. 
					
\begin{normalbox}
\textbf{Preliminary Study:}
Almost all participants (98\%) stated that they use SO for both private and work-related projects.
Half of them (49\%) reported that the last time they copied or adapted a code snippet from SO, they did not attribute its origin; 40\% added a source code comment with a link to the corresponding question or answer.
\end{normalbox}

\section{Usage Without Attribution (RQ1 -- Phase 1)}
\label{sec:phase1}

In our preliminary study, many developers reported that they did not attribute code snippets copied from SO.
Most participants who did attribute the snippets added a source code comment with a link to the corresponding question or answer.
Thus, we decided to utilize BigQuery to find all links to SO questions and answers in all Java files in the GitHub data set.
Afterwards, we built regular expressions matching the snippets from the ten most frequently referenced Java answers and searched for matches in all Java files in the data set to detect unattributed usages of those snippets.

\subsection{Method:}

\begin{figure}
\centering
\includegraphics[width=1\columnwidth,  trim=0.0in 0.0in 0.0in 0.0in]{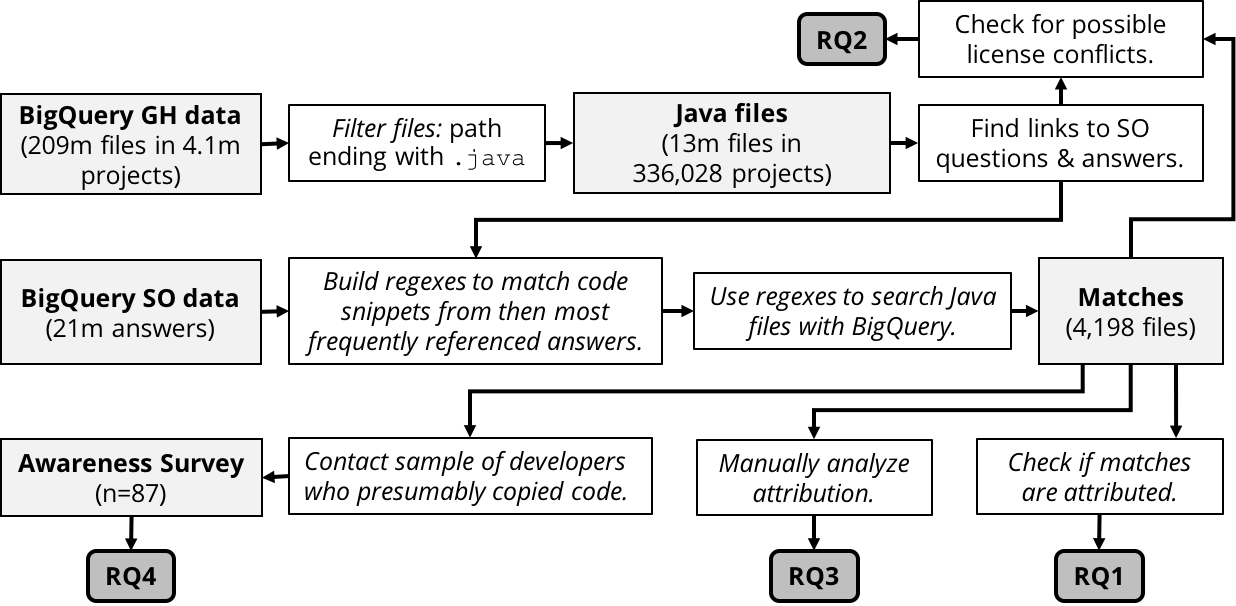}
\caption{RQ1-4 -- Phase 1: We searched for attributed and unattributed usages of the code snippets from the ten most frequently referenced answers on Stack Overflow (SO) in all Java files in the BigQuery GitHub (GH) data set using regular expressions and used this data to answer our research questions (time span of this phase: 07/2016--11/2016).}
\label{fig:phase1}
\end{figure}

Figure~\ref{fig:phase1} visualizes our initial workflow for finding attributed and unattributed usages (including the connection to other research questions).
We considered all files ending with \texttt{.java} to be Java source code files and applied the following regular expression (regex) to each line of those files:

\begin{regex}
(?i:https?://stackoverflow\.com/[^\s)\.\"]*)
\end{regex}

Because there are different ways of referring to questions and answers on SO, i.e., using full URLs or short URLs, we mapped all extracted URLs to their corresponding sharing link (ending with \verb+/q/<id>+ for questions and \verb+/a/<id>+ for answers). In the following, we use the term \textit{reference} to denote a link to content on SO.
In the database schema of the \textit{BigQuery GH data set}, copied files have the same ID (hash value of the content).
For our analysis, we only considered distinct references, meaning that we counted references in files with the same content only once.
Because many files on GitHub are duplicates~\citep{GharehyazieRayOthers2017, LopesMajOthers2017}, we distinguish between the number of \textit{distinct referencing files}, meaning the number of distinct files in which a URL was present in a source code comment, and the number of \textit{distinct referencing lines}, meaning the number of distinct source code lines in which a URL was used (exact string match including whitespaces).
The former may exaggerate the number of distinct references as files may be copied and then slightly changed, the latter may understate the number of distinct references as two developers may independently use the same source code line to reference a question or an answer.
According to Google, most forks were excluded in their \textit{BigQuery GH data set}. %~\citep{Hoffa16}.
In the first phase, we relied on the unique file IDs to exclude copied files.
In the third phase, we further excluded all repositories that were marked as forks in the \textit{BigQuery GHTorrent data set} (see Section~\ref{sec:phase3}).

Our first approach for finding unattributed usages of SO snippets utilized manually created regular expressions to find matches of non-trivial code snippets in all public GH projects containing Java code.
Since this approach is time-consuming, we had to carefully select the snippets for which we then built the regular expressions.
We decided to extract the code snippets from the ten most frequently referenced Java answers on SO, because we thought that these snippets are likely to be also used without attribution (assuming that the attribution ratio is relatively stable across posts).
In a next step, we randomly chose (up to) ten Java source code files referencing the corresponding SO answer.
Then, we manually created a regex for each SO snippet % using the online regular expression tester \texttt{regex101.com}.
and iteratively modified it to match both the snippet and as many of the referencing Java files as possible, while taking care that it does not become too generic, leading to false positives.

Table~\ref{tab:java-top10-1} lists the ten most frequently referenced Java answers.
In the table, we included a short description of the thread's topic and mention whether the code in the answer is a whole class, a single method, or just a few lines of code (snippet).
We also added information about the external availability of the source code from the SO post.
The top-ranked Java snippet was also available on a personal blog post by the same author.
However, the author has the copyright for his blog post and provides no license, thus only the SO post allows the usage of that snippet.
Further, the SO thread is the first result on Google (as of June 8, 2017) when searching for ``human readable byte size java''.
Therefore, the SO post is likely the primary source for copying this particular snippet.
The second snippet is based on a blog post by a different author, also copyrighted without providing a license.
Moreover, this blog post is only available using the \textit{Internet Archive Wayback Machine}.\footnote{\url{http://web.archive.org/}}
Therefore, also for this snippet SO is likely to be the primary source.
The third snippet is based on a different SO post, but had been adapted. Thus, the license is still CC BY-SA 3.0.
For the other snippets, we could not identify an external source with a different license.

\begin{table}
\caption{RQ 1 -- Phase 1: Ten most frequently referenced code snippets from SO Java answers; one asterisk: link was broken and referred to a question, we selected two referenced snippets; two asterisks: snippet based on external resource, but adapted.}
\begin{tabular}{cllll}
\hline\noalign{\smallskip}
Rank & Answer ID & Description & Type & Ext. Availability\\
\noalign{\smallskip}\hline\noalign{\smallskip}
1 & \href{http://stackoverflow.com/a/3758880}{3758880} & human readable byte size & method & blog, no license\\
2 & \href{http://stackoverflow.com/a/5445161}{5445161**} & read InputStream to String & method & blog, no license\\
3 & \href{http://stackoverflow.com/a/9855338}{9855338**} & convert byte array to hex String & method & other SO post\\
4 & \href{http://stackoverflow.com/a/26196831}{26196831} & Android: RecyclerView onClick & class & none\\
5 & \href{http://stackoverflow.com/a/7696791}{7696791}* & Android: close soft keyboard & snippet & none\\
6 & \href{http://stackoverflow.com/a/140861}{140861} & hex dump String to byte array & class & none\\
7 & \href{http://stackoverflow.com/a/2581754}{2581754} & sort Map$<$Key, Value$>$ by values & class & none\\
8 & \href{http://stackoverflow.com/a/5599842}{5599842} & format file size as MB, GB, etc. & method & none\\
9 & \href{http://stackoverflow.com/a/326440}{326440} & create Java String from file cont. & method & none\\
10 & \href{http://stackoverflow.com/a/3145655}{3145655} & Android: get current location & class & none\\
\noalign{\smallskip}\hline
\end{tabular}
\label{tab:java-top10-1}
\end{table}

Table~\ref{tab:java-top10-2} shows, for each of the ten Java answers, the number of distinct referencing lines ($\textsc{l}_\textsc{a}$) and the number of distinct referencing files ($\textsc{f}_\textsc{a}$).
Further, we provide the number of distinct files with a reference to either to the answer or to the corresponding question ($\textsc{f}_\textsc{aq}$).
For this value we do not know if the developer actually wanted to refer to the snippet from the specific answer we are considering or to another answer from the same thread.
The table also shows the number of GH references we used to test the regular expression and how many of those references the regex matched.

We used BigQuery's \texttt{REGEXP\_MATCH} function to check all Java files in the GH data set for matches of each regex.
We provide the extracted SO snippets, the referencing Java code from GH, the regular expressions, and the SQL scripts as supplementary material~\citep{Baltes2018e}.

\subsection{Results:}
\label{sec:phase1-results}

\begin{table}
\caption{RQ 1 -- Phase 1: Ten most frequently referenced code snippets from SO Java answers, references in GH Java files and testing of regular expressions for those snippets; $\textsc{l}_\textsc{a}$: number of distinct referencing lines, $\textsc{f}_\textsc{a}$: number of distinct referencing files, $\textsc{f}_\textsc{aq}$: number of distinct referencing files including references to corresponding question.}
\begin{tabular}{crrrrrr}
\hline\noalign{\smallskip}
Rank & \multicolumn{3}{c}{References} & \multicolumn{3}{c}{Regex}\\
& $\textsc{l}_\textsc{a}$ & $\textsc{f}_\textsc{a}$ & $\textsc{f}_\textsc{aq}$ & \textsc{tested} & \textsc{matched} & \textsc{recall} \\
\noalign{\smallskip}\hline\noalign{\smallskip}
1 & 21 & 43 & 122 & 10 & 9 & 90.0\%  \\
2 & 20 & 39 & 100 & 10 & 7 & 70.0\% \\
3 & 19 & 27 & 108 & 10 & 10 & 100.0\% \\
4 & 12 & 15 & 19 & 10 & 9 & 90.0\% \\
5 & 9 & 20 & 34 & 9 & 4 & 44.4\% \\
6 & 8 & 12 & 74 & 7 & 7 & 100.0\% \\
7 & 8 & 9 & 41 & 8 & 8 & 100.0\% \\
8 & 7 & 17 & 36 & 7 & 5 & 71.4\% \\
9 & 7 & 12 & 47 & 7 & 1 & 14.3\% \\
10 & 7 & 12 & 26 & 6 & 6 & 100.0\% \\
\noalign{\smallskip}\hline\noalign{\smallskip}
All & 118 & 206 & 607 & 84 & 66 & $M$ 78.0\% \\
\noalign{\smallskip}\hline
\end{tabular}
\label{tab:java-top10-2}
\end{table}

Table~\ref{tab:java-top10-3} shows how many files of the data set each regex matched and how many of those matches were distinct files.
We report how many of the matched files contained a reference to the answer or the corresponding question (\textsc{ref}) and how many files did not contain a reference (\textsc{no-ref}).
We also calculated the recall by comparing $\textsc{f}_\textsc{aq}$ and \textsc{ref}, i.e., the number of distinct files with a reference to either the answers or the corresponding question and the number of matched files containing such a reference.
This allowed us to assess how good the regex was in matching possible duplicates of the snippet.

We calculated two estimates for the ratio of files with attributed snippets:
First, we compared the number of distinct referenced matches (\textsc{ref}) to the total number of distinct matches (\textsc{distinct}).
The second estimate is the number of distinct matches with a reference either to the answer or to the corresponding question ($\textsc{f}_\textsc{aq}$) compared to the number of distinct matches (\textsc{distinct}).
%We set the result to the minimum of the ratio and 100\% in case \textsc{no-ref} was smaller than $\textsc{f}_\textsc{aq}$ (see snippet 10).
Please note that the comparisons with $\textsc{f}_\textsc{aq}$ understate the recall and overstate the attribution ratio, because $\textsc{f}_\textsc{aq}$ likely includes references to other answers of the thread.
To evaluate the number of false positives, we checked (up to) 50 matches for each regex and found no match that we did not consider to be a clear copy of the snippet.

\begin{table}
\caption{RQ 1 -- Phase 1: Ten most frequently referenced code snippets from SO Java answers; estimated ratio of unattributed usages detected using regular expressions; number of matched files (\textsc{all}), distinct matches (\textsc{distinct}), distinct matches with reference to SO (\textsc{ref}), distinct matches without reference to SO (\textsc{no-ref}).}
\begin{tabular}{crrrrrrr}
\hline\noalign{\smallskip}
Rank & \multicolumn{4}{c}{Matches} & \multicolumn{1}{c}{Recall} & \multicolumn{2}{c}{Attribution} \\
& \textsc{all} & \textsc{distinct} & \textsc{ref} & \textsc{no-ref} & $\textsc{ref} / \textsc{f}_\textsc{aq}$ & $\textsc{ref} / \textsc{distinct}$ & $\textsc{f}_\textsc{aq} / \textsc{dist.}$ \\
\noalign{\smallskip}\hline\noalign{\smallskip}
1 & 997 & 448 & 97 & 351 & 79.5\% & 21.7\% & 27.2\% \\
2 & 1,843 & 913 & 60 & 853 & 60.0\% & 6.6\% & 11.0\% \\
3 & 2,662 & 902 & 87 & 815 & 80.6\% & 9.6\% & 12.0\%\\
4 & 420 & 170 & 18 & 152 & 94.7\% & 10.6\% & 11.2\% \\
5 & 1,492 & 402 & 25 & 377 & 73.5\% & 6.2\% & 8.5\% \\
6 & 2,642 & 807 & 65 & 742 & 87.8\% & 8.1\% & 9.2\% \\
7 & 160 & 124 & 12 & 112 & 29.3\% & 9.7\% & 33.1\% \\
8 & 355 & 174 & 22 & 152 & 61.1\% & 12.6\% & 20.7\% \\
9 & 295 & 225 & 5 & 220 & 10.6\% & 2.2\% & 20.9\% \\
10 & 65 & 33 & 11 & 22 & 42.3\% & 33.3\% & 78.8\% \\
\noalign{\smallskip}\hline\noalign{\smallskip}
All & 10,931 & 4,198 & 402 & 3,796 & $M$ 61.9\% & $M$ 12.1\% & $M$ 23.2\% \\
\noalign{\smallskip}\hline
\end{tabular}
\label{tab:java-top10-3}
\end{table}

To illustrate the procedure, we present the snippet from the most frequently referenced Java answer and the corresponding regex below.
The snippet is a method returning a human-readable string representation of a byte value (e.g., for 1024 it returns 1.0 kB or 1.0 KiB)~\citep{Aioobe2010}.
It was referenced in 21 distinct lines and in 43 distinct files, meaning that several files used the same line content to reference the snippet.
Together with the corresponding question, we found 122 distinct referencing files (see Table~\ref{tab:java-top10-2}).

% public static
\begin{java}
String humanReadableByteCount(long bytes, boolean si) {
    int unit = si ? 1000 : 1024;
    if (bytes < unit) return bytes + " B";
    int exp = (int) (Math.log(bytes) / Math.log(unit));
    String pre = (si ? "kMGTPE" : "KMGTPE").charAt(exp-1) + (si ? "" : "i");
    return String.format("%.1f %sB", bytes / Math.pow(unit, exp), pre); }
\end{java}

Starting with the above snippet, we created a regular expression and iteratively refined it until it matched 9 out of 10 referencing files from GH.
%An excerpt of the final regular expression can be found below.
The final regular expression, which can be found below, matched 80\% of the files containing a reference either to the answer itself or the corresponding question.

\begin{regex}
((?i:String[\s]+\w+\([^\{]*long[^\{]+\)[\s]*\{[\s\S]+if
[\s]*\([^<]+<[^\)]+\)[\s\S]*return[^;]+\+[^;]*\"\ B\"
[\s\S]+int[\s][^\=]+\=[\s]*\([\s]*int[\s]*\)[\s]*\([\s]*
Math[\s]*\.[\s]*log[\s]*\([^\)]+\)[\s]*\/[\s]*Math[\s]*
\.[\s]*log[\s]*\([^\)]+\)[\s]*\)[\s\S]+return[^\}]+
String[\s]*\.[\s]*format[\s]*\([^\}]+\}))
\end{regex}
%\begin{regex}
%((?i:String[\s]+\w+\([^\{]*long[^\{]+\)[\s]*\{[\s\S]+if 
%... log[\s]*\([^\)]+\)[\s]*\)[\s\S]+return[^\}]+
%String[\s]*\.[\s]*format[\s]*\([^\}]+\}))
%\end{regex}

Only 21.7\% of the matched files were attributed; compared to $\textsc{f}_\textsc{aq}$ the ratio was 27.2\%.
On average, the regular expressions we created matched 78.0\% of the referencing GH Java files from the test sets (see Table~\ref{tab:java-top10-2}) and 61.9\% of the files in $\textsc{f}_\textsc{aq}$ (see Table~\ref{tab:java-top10-3}).
The average ratio of attributed matches was 12.1\%; compared to $\textsc{f}_\textsc{aq}$, the ratio was still only 23.2\%.
As motivated above, the latter ratio overstates the amount of referenced usages and can thus be considered an upper bound.
Because the regular expressions were rather strict and false positives were not present in the samples we checked, we can estimate that at most 23.2\% of the copies of the ten most frequently referenced SO Java code snippets are being attributed when copied into Java files on GH.

\begin{normalbox}
\textbf{Usage Without Attribution (RQ1 -- Phase 1):}
At most 23.2\% of the copies of code snippets from the ten most frequently referenced SO Java answers in Java files on GH were attributed using a link to SO.
\end{normalbox}

\section{Usage Without Attribution (RQ1 -- Phase 2)}
\label{sec:phase2}

To triangulate our estimate from the first phase that at most 23.2\% of the usages of SO code snippets in GH projects are attributed, we followed a second approach and used a token-based code clone detector, the \textit{PMD Copy-Paste Detector} version 5.4.1~\citep{PMD2016}, to find unreferenced usages of SO code snippets in a random sample of popular GH Java projects.

%Too small and trivial snippets (e.g., simple API calls or event handlers) may lead to many false positive results. Also, no maintainability or legal issues arise from the presence of such trivial code snippets in public GH projects.

\subsection{Method:}

We decided to use the \textit{PMD Copy-Paste Detector} (CPD) for finding clones of SO snippets, because this tool is open-source, actively developed, and widely used. It is integrated into the IntelliJ Java IDE %~\citep{JetBrains17}
and there are plugins for other IDEs as well.  

The detection of code clones within a set of source code files is a computationally expensive task. Therefore, we had to restrict our analysis to a sample of GH Java projects and a sample of Java code snippets from SO.
A random sample of GH projects would contain many small personal projects, homework assignments, or other projects that are not ``engineered software projects''~\citep{KalliamvakouGousiosOthers2014, MunaiahKrohOthers2017}.
Filtering projects according to their popularity, measured using the number of watchers or stargazers, has been used in several well-received studies and proved to have a very high precision (almost 0\% false positives)~\citep{MunaiahKrohOthers2017}.
Hence we applied a similar filtering strategy.

\begin{figure}
\centering
\includegraphics[width=0.9\columnwidth,  trim=0.05in 0.2in 0.55in 0.2in]{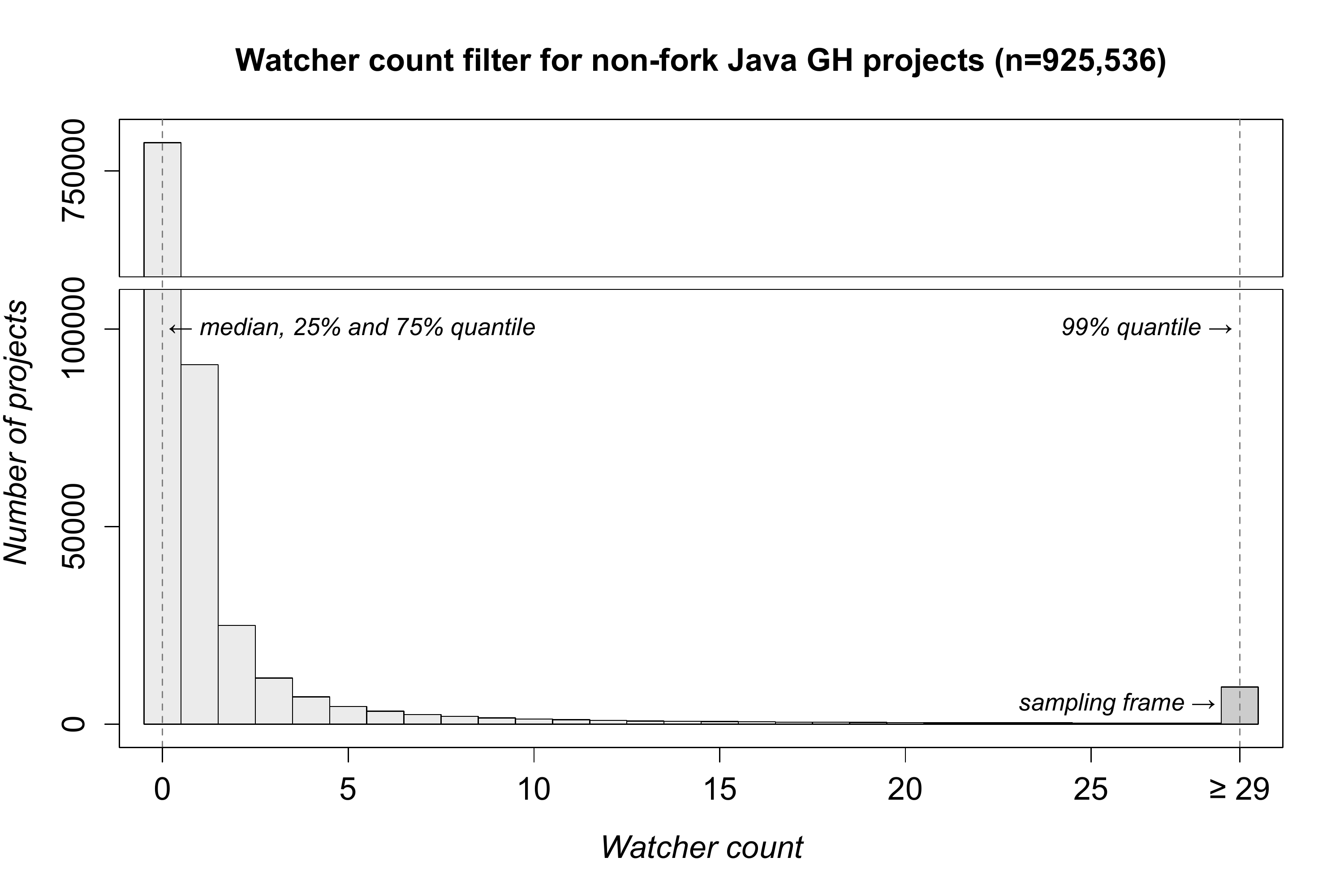} % left bottom right top
\caption{RQ1 -- Phase 2: Histogram visualizing the selected sampling frame of popular GitHub Java projects ($n=9,437$); the 99\% quantile of all non-fork Java projects was 29 watchers ($M=2.77$, $Mdn=0$, $Q_{1,3}=0$); based on the GHTorrent data dump 2016-02-01.}
\label{fig:phase2-filter-histogram}
\end{figure}

Our sampling frame consisted of all Java projects in the \textit{GHTorrent data set} (February 16, 2016) that were no forks, not deleted, and had at least 29 watchers (99\% quantile for all Java projects, see Figure~\ref{fig:phase2-filter-histogram}).
We excluded forks, because they may skew the results by adding almost identical repositories to the sample.
We excluded deleted repositories, because we would not be able to retrieve the source code of such repositories.
From this sampling frame ($n=9,437$), we randomly selected 3,000 Java projects.
We were able to successfully download 2,313 of them.
Some downloads failed, because our script only tried to retrieve the master branch, which may not exist, and some repositories may have been renamed or deleted between the creation of the \emph{GHTorrent data dump} and the time we downloaded the sample (April 21, 2016).

We searched for two different sets of SO code snippets in the sample of GH projects:
One set with snippets that had referenced usages in the GH projects under analysis ($S_\text{gh}$) and one set with popular Java snippets identified using the data from the first phase of our research ($S_\text{top100}$).
The first set allowed us to compare referenced usages of SO snippets to unreferenced usages of the same snippets; the second set allowed us to analyze how many copies of popular SO Java snippets were being attributed in the sample of GH projects.
For the first set, we searched for references to SO in all Java files in the project sample using the same regular expression as in the first phase.
We then manually extracted the snippets from all referenced answers, dropping answers that did not contain code or only trivial snippets (e.g., simple API calls, snippets for conceptual questions, etc.).
This resulted in a total number of 137 extracted SO snippets.
For the second set of snippets, we manually extracted the code from the 100 most frequently referenced Java answers, identified using the same data and ranking approach as in the first phase (see Section~\ref{sec:phase1}).
We used the number of distinct referencing lines as the primary and the number of distinct referencing files as the secondary sort key.
This resulted in 111 snippets.
We provide all extracted snippets and the names of all analyzed Java projects as supplementary material~\citep{Baltes2018e}.

As a last preparation step, we checked the intersection of the two snippet sets to prevent snippets that are in both sets biasing the results.
We identified 26 snippets from 18 answers to be in $S_\text{gh} \cap S_{top100}$.
We present the results for each snippet set separately and count the intersecting snippets and matches only once in the summary.
Before presenting the results, we describe how we calibrated CPD for finding SO snippets in Java projects.

\subsection{Calibration of the Code Clone Detector:}

We iteratively optimized CPD's parameters using $S_\text{gh}$ as ground truth, because for this set of snippets we already identified the attributed usages and could thus determine precision and recall.
For Java files, the relevant parameters to configure CPD are the minimum token length that should be reported as a duplicate (\texttt{mt}) % (-{}-minimum-tokens);
and three boolean flags to configure text comparison: One to ignore language annotations (\texttt{ia}), % (-{}-ignore-annotations); 
one to ignore constant and variable names (\texttt{ii}), % (-{}-ignore-identifiers);
and one to ignore number values and string contents (\texttt{il}). % (-{}-ignore-literals).
To compare the results of different iterations, we used the following definitions of \textit{precision} and \textit{recall}:

\begin{definition}
Let $C$ (copies) be a relation over a set of code snippets $S$ and a set of source code files $F$:
\[ C \subseteq S \times F \]
Let $C_\text{so} \subseteq C$ be the set of copies identified by an SO answer URL in the source code file and
$C_\text{cpd} \subseteq C$ be the set of copies identified by CPD. Then we define precision and recall as follows:
\[ \text{precision} = \frac{ | C_\text{so} \cap C_\text{cpd} | }{ | C_\text{cpd} | } \hspace{4em} \text{recall} = \frac{ | C_\text{so} \cap C_\text{cpd} | }{ | C_\text{so} | } \]
\end{definition}

Please note that the precision may be $<\!1$ even if all copies found by CPD are actually duplicates of a snippet in $S_{so}$.
The reason for this is that the Java files in our test set may contain copies of these snippets that are either unreferenced or are referenced using a link to the question. % instead of the answer.
As CPD cannot be configured to only find clones of one set of files in another, we wrote a wrapper to exclude matches within the snippets and within the analyzed projects.
The wrapper returns the matches between snippets and Java files in the projects along with the line numbers of the exact positions of each match.
From this data, we derived the relation $C_\text{cpd}$.
An example for one entry is provided below:
\begin{code}
so-answer-3054692, Floobits-floobits-intellij/.../Utils.java
\end{code}
In this example, the snippet extracted from the SO answer with ID \texttt{3054692} was found in the file identified by the given path (the root is the name of the GH repository).

We derived $C_\text{so}$ from the references we already extracted ($S_\text{gh}$).
Using these two relations, we calculated precision and recall for each test run according to the above definitions. 

\begin{figure}
\centering
\includegraphics[width=0.9\columnwidth,  trim=0.0in 0.2in 0.0in 0.2in]{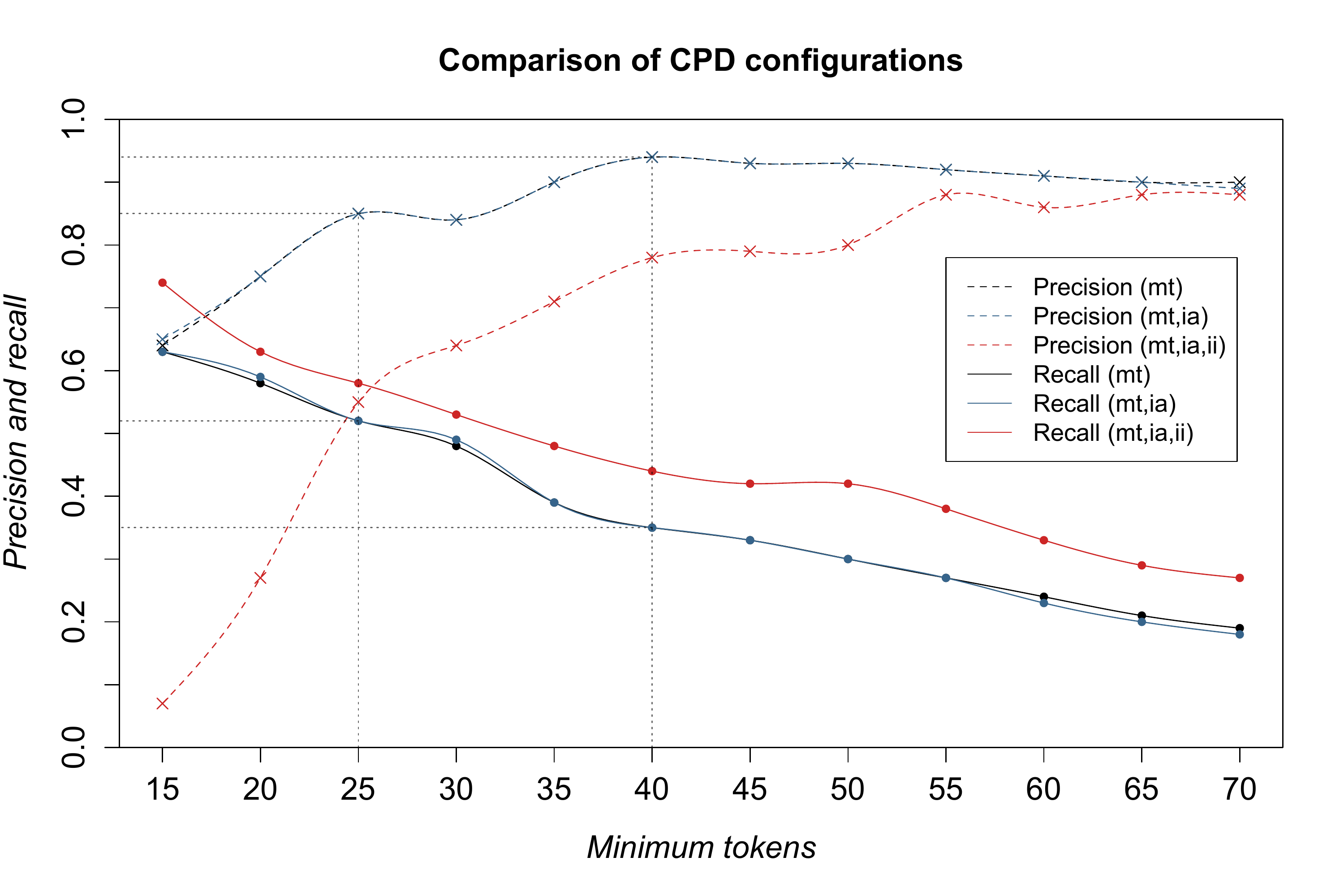} % left bottom right top
\caption{RQ1 -- Phase 2: Comparison of different CPD configurations: black: only \texttt{mt} set; blue: \texttt{mt} and \texttt{ia} set; red: \texttt{mt}, \texttt{ia}, and \texttt{ii} set; dashed line: precision, solid line: recall; final configuration: $\texttt{mt} = 40$ ($\text{precision} = 0.94$, $\text{recall} = 0.35$).}
\label{fig:precision-recall}
\end{figure}

Figure~\ref{fig:precision-recall} shows the results for different configurations of CPD.
We conducted three test runs with $\texttt{mt} \in \{ 15, 20, ..., 95, 100 \}$: (1) without further parameters, (2) with flag \texttt{ia} set, and (3) with flags \texttt{ia} and \texttt{ii} set.
First, we also included the flag \texttt{il}, but with the relatively small values we used for \texttt{mt} this resulted in too many false positive results.
Moreover, setting \texttt{il} lead to a runtime that was magnitudes longer than the other configurations.
Because our goal was to increase precision and avoid false positives, we dropped \texttt{ii} despite the slightly higher recall.
Since the flag \texttt{ia} had almost no effect on precision and recall (only few snippets with annotations in $S_\text{gh}$), we also dropped it.

\begin{table}
\caption{RQ 1 -- Phase 2: Results for different CPD configurations; all matches, distinct snippet-file pairs, true positive matches ($C_\text{so}\,\cap\,C_\text{cpd}$), false positive matches ($C_\text{cpd}\!\setminus\!C_\text{so}$), precision, and recall.}
\begin{tabular}{crrrrrr}
\hline\noalign{\smallskip}
\multicolumn{1}{c}{Configuration} & \multicolumn{6}{c}{Matches} \\
& \multicolumn{1}{c}{\textsc{all}} & \multicolumn{1}{c}{\textsc{distinct}} & \multicolumn{1}{c}{\textsc{True pos.}} & \multicolumn{1}{c}{\textsc{False pos.}} & \multicolumn{1}{c}{\textsc{Precision}} & \multicolumn{1}{c}{\textsc{Recall}} \\
\noalign{\smallskip}\hline\noalign{\smallskip}
$\texttt{mt}{=}40$ & $103$ & $51$ & $48$ & $3$ & $94\%$ & $35\%$ \\
$\texttt{mt}{=}25$ & $268$ & $84$ & $72$ & $12$ & $85\%$ & $53\%$ \\
\noalign{\smallskip}\hline
\end{tabular}
\label{tab:cpd-config-results}
\end{table}

We achieved the highest precision by setting $\texttt{mt}{=}40$ without further parameters ($\textit{prec}{=}0.94$, $\textit{rec}{=}0.35$).
We selected $\texttt{mt}{=}25$ as a second candidate because of its higher recall ($\textit{prec}{=}0.85$, $\textit{rec}{=}0.53$).
Table~\ref{tab:cpd-config-results} shows the results for these two configurations.
We divided the matched files into \textit{true positive} ($C_\text{so}\,\cap\,C_\text{cpd}$) and \textit{false positive} results ($C_\text{cpd}\!\setminus\!C_\text{so}$).
We manually investigated all true and false positives for the two configurations and found that all matches were true positives; the false positives were clones that were not referenced.
Nevertheless, the configuration with $\texttt{mt}{=}25$ contained some relatively small matches, e.g., parts of for-loops, that were likely to produce false positives outside of our test collection.
Based on the results of our test runs, we chose $\texttt{mt}{=}40$ for the final CPD configuration.
As we decided not to set \texttt{ia} and \texttt{ii}, this configuration can only detect type-1 clones of the snippets, i.e., copied code that only varies in whitespace, layout, or comments~\citep{RoyCordyOthers2009}.
%We provide all scripts and data from the test runs as supplementary material~\citep{Baltes2018e}.

%\todo{get token count for snippets in samples}
%\todo{Edit-Distanz für gematchte snippets mit original berechnen?}

\subsection{Results:}

Using the configuration $\texttt{mt}{=}40$, we searched for type-1 clones of the two snippet sets $S_\text{gh}$ and $S_\text{top100}$ in two separate runs.
Each run took between 8 and 9 hours on a regular Desktop PC running Ubuntu 14.04 LTS (Intel Core i5-4670, 16 GB RAM, SSD).
Table~\ref{tab:snippet-sets} lists the results for each snippet set.
It shows the number of snippets in each set, the number of answers from which the snippets were extracted, and the number of matched snippets, answers, and files.
In the analyzed GH projects, we found 634 Java files from 274 projects that contained a reference to SO (0.14\% of all Java files in the sample and 12\% of all projects.
The table shows how many of the matched files contained a reference to SO and the number of repositories containing a matched file.
%We did not check whether the references in the files were in fact for the found snippet or another part of the source code.

In a first data cleaning step, we analyzed the results and found that one of the snippets in $S_\text{top100}$ was responsible for 272 matches (48\% of all matches).
This snippet contained a long list of invalid characters in file names.
We looked at the matched files and found that most of the matches used this array in another context than the SO snippet.
Thus, we considered these matches to be false positives and excluded them from our analysis.
To estimate the number of false positives in the remaining matches, we randomly chose 100 distinct matches (snippet-file pairs) from each set and manually checked whether the files actually contained a copy of the snippet.
We rated all analyzed matches as true positives.

We further checked if the snippets were available from an external source, meaning a website, blog, or source code repository outside of the SO platform.
If snippets were also available outside of SO, more permissive licenses could apply that allow using the snippet without attributing SO as the source.
We followed all links in the SO answers from which the snippets were taken and checked if the snippet was available in the linked resource.
If it was available, we searched the websites for licenses or terms of service that apply for the content.
Tables~\ref{tab:snippet-sets-external-source}~and~\ref{tab:snippet-sets-license} summarize the results of this analysis.
Table~\ref{tab:snippet-sets-external-source} shows how many answers provided an external source for the snippet (12\%), together with the type of the source.
We found copies of the snippets in blog posts (8), GitHub repos (6), Android or Java bug reports (5), and in the official Android or Java documentation (2).
For the answers having an external source, Table~\ref{tab:snippet-sets-license} shows if this source allows to use the snippet under a more permissive license than CC BY-SA 3.0.
Twelve of those 21 answers provided a license or terms of service, of which only three were more permissive than Stack Overflow's license:
In one case,\footnote{\url{http://balusc.omnifaces.org/2007/07/fileservlet.html}} the author added a comment indicating that the snippet is free to use: ``There is no copyright on the code. You can copy, change and distribute it freely. Just mentioning this site should be fair''; two sources were licensed under the Apache 2.0 license.
One source was licensed under the GNU GPL 2.0, which is also a copyleft license and hence not more permissive than CC BY-SA 3.0; the other eight sources had terms of service restricting the usage of the snippet.
We can conclude that even if some snippets are also available outside of SO, this does not necessarily mean that the external sources are more permissive than SO's license.

\begin{table}
\caption{RQ 1 -- Phase 2: Results of searching copies of two sets of Stack Overflow snippets in a sample of GitHub projects ($n=2,313$): Columns named \textsc{matched} show number of distinct matched snippets, answers, files, and repos; column \textsc{ref} shows number of matched files containing a reference to Stack Overflow.}
\begin{tabular}{lrrrrrrr}
\hline\noalign{\smallskip}
\multicolumn{1}{c}{\multirow{2}{*}{Set}} & \multicolumn{4}{c}{Snippets} & \multicolumn{2}{c}{Files} & \multicolumn{1}{c}{Repos}\\
& \multicolumn{1}{c}{\textsc{all}} & \multicolumn{1}{c}{\textsc{matched}} & \multicolumn{1}{c}{\textsc{answers}} & \multicolumn{1}{c}{\textsc{matched}} & \multicolumn{1}{c}{\textsc{match.}} & \multicolumn{1}{c}{\textsc{ref}} & \multicolumn{1}{c}{\textsc{matched}} \\
\noalign{\smallskip}\hline\noalign{\smallskip}
$S_\text{gh}$ & 137 & 53 (39\%) & 102 & 52 (51\%) & 163 & 58 (36\%) & 124 (5\%) \\
$S_\text{top100}$ & 111 & 48 (43\%) & 85 & 46 (54\%) & 173 & 25 (14\%) & 125 (5\%) \\
\noalign{\smallskip}\hline\noalign{\smallskip}
$\cup S$ & 222 & 101 (46\%)  & 169 & 86 (51\%) & 297 & 70 (24\%) & 199 (9\%) \\
\noalign{\smallskip}\hline
\end{tabular}
\label{tab:snippet-sets}
\end{table}

\begin{table}
\caption{RQ 1 -- Phase 2: External sources for snippets: The table shows the number of answers with snippets in the two sets and how many of those answers contained a link to an external source.
Abbreviations: Snippets also available in a blog post (\textsc{blog}), in a GitHub repository (\textsc{GH}), in an Android or JDK bug description (\textsc{bug report}), in an Android or Java documentation page (\textsc{doc}).}
\begin{tabular}{lrrrrrrrr}
\hline\noalign{\smallskip}
\multicolumn{1}{c}{\multirow{2}{*}{Set}} & \multicolumn{6}{c}{External source in SO answers} \\
 & \multicolumn{1}{c}{\textsc{all}} & \multicolumn{1}{c}{\textsc{no}} & \multicolumn{1}{c}{\textsc{yes}} & \multicolumn{1}{c}{\textsc{blog}} & \multicolumn{1}{c}{\textsc{GH}} & \multicolumn{1}{c}{\textsc{Bug report}} & \multicolumn{1}{c}{\textsc{doc}} \\
\noalign{\smallskip}\hline\noalign{\smallskip}
$S_\text{gh}$ & 102 & 89 (87\%) & 13 (13\%) & 6 (6\%) & 2 (2\%) & 4 (4\%) & 1 (1\%) \\
$S_\text{top100}$ & 85 & 76 (89\%) & 9 (11\%) & 2 (2\%) & 5 (6\%) & 1 (1\%) & 1 (1\%) \\
\noalign{\smallskip}\hline\noalign{\smallskip}
$\cup S$ & 169 & 148 (88\%) & 21 (12\%) & 8 (5\%) & 6 (4\%) & 5 (3\%) & 2 (1\%) \\
\noalign{\smallskip}\hline
\end{tabular}
\label{tab:snippet-sets-external-source}
\end{table}

\begin{table}
\caption{RQ 1 -- Phase 2: License of external sources for snippets: The table shows under which licenses the snippets from external sources can be used; \textsc{no}: no license provided, \textsc{free}: author added a comment that the code is free to use, \textsc{ToS}: usage is restricted by the website's terms of service, \textsc{Apache 2.0}: available under the Apache 2.0 license, \textsc{GPL 2.0}: available under the GPL 2.0 license.}
\begin{tabular}{lrrrrrrrr}
\hline\noalign{\smallskip}
\multicolumn{1}{c}{\multirow{2}{*}{Set}} & \multicolumn{7}{c}{License of external sources} \\
 & \multicolumn{1}{c}{\textsc{all}} & \multicolumn{1}{c}{\textsc{no}} & \multicolumn{1}{c}{\textsc{yes}} & \multicolumn{1}{c}{\textsc{ToS}} & \multicolumn{1}{c}{\textsc{free}} & \multicolumn{1}{c}{\textsc{Apache 2.0}} & \multicolumn{1}{c}{\textsc{GPL 2.0}} \\
\noalign{\smallskip}\hline\noalign{\smallskip}
$S_\text{gh}$ & 13 & 4 (31\%) & 9 (69\%) & 7 & 1 & 1 & 0 \\
$S_\text{top100}$ & 9 & 6 (67\%) & 3 (33\%) & 1 & 0 & 1 & 1 \\
\noalign{\smallskip}\hline\noalign{\smallskip}
$\cup S$ & 21 & 9 (43\%) & 12 (57\%) & 8 & 1 & 2 & 1 \\
\noalign{\smallskip}\hline
\end{tabular}
\label{tab:snippet-sets-license}
\end{table}

Overall, CPD found one or more copies of snippets from the two snippet sets in 297 distinct files. % (0.07\% of all files in the sample).
The identified clones were duplicates of 101 different snippets (46\% of all distinct snippets in the sets) from 86 answers (51\% of all answers in the sets).
Only 70 matched files (24\%) contained a reference to a SO question or answer.
In summary, 199 repositories (9\% of all repositories in the sample) contained files with copies of snippets from SO.
As we did not observe any false positive results (except for the match we excluded in the data cleaning step, see above), the number of matches can be interpreted as a lower bound for the amount of copies that are actually present in the sample.
%Moreover, we considered files as having a reference to SO if they contained a SO URL, even if this URL was a reference to another thread that had nothing to do with the snippet.
%Thus, the estimates for the amount of referenced snippets can be interpreted as an upper bound for the actual ratio.

\begin{normalbox}
\textbf{Usage Without Attribution (RQ1 -- Phase 2):}
Using CPD, we found that in a sample of popular Java projects ($n{=}2,313$), 199 repositories (9\%) contained a copy of one of the 222 SO snippets we considered.
Only 24\% of the matched files contained a reference to SO as required by SO's license.
\end{normalbox}

\section{Usage Without Attribution (RQ1 -- Phase 3)}
\label{sec:phase3}

\begin{figure}
\centering
\includegraphics[width=0.8\columnwidth,  trim=0.0in 0.0in 0.0in 0.0in]{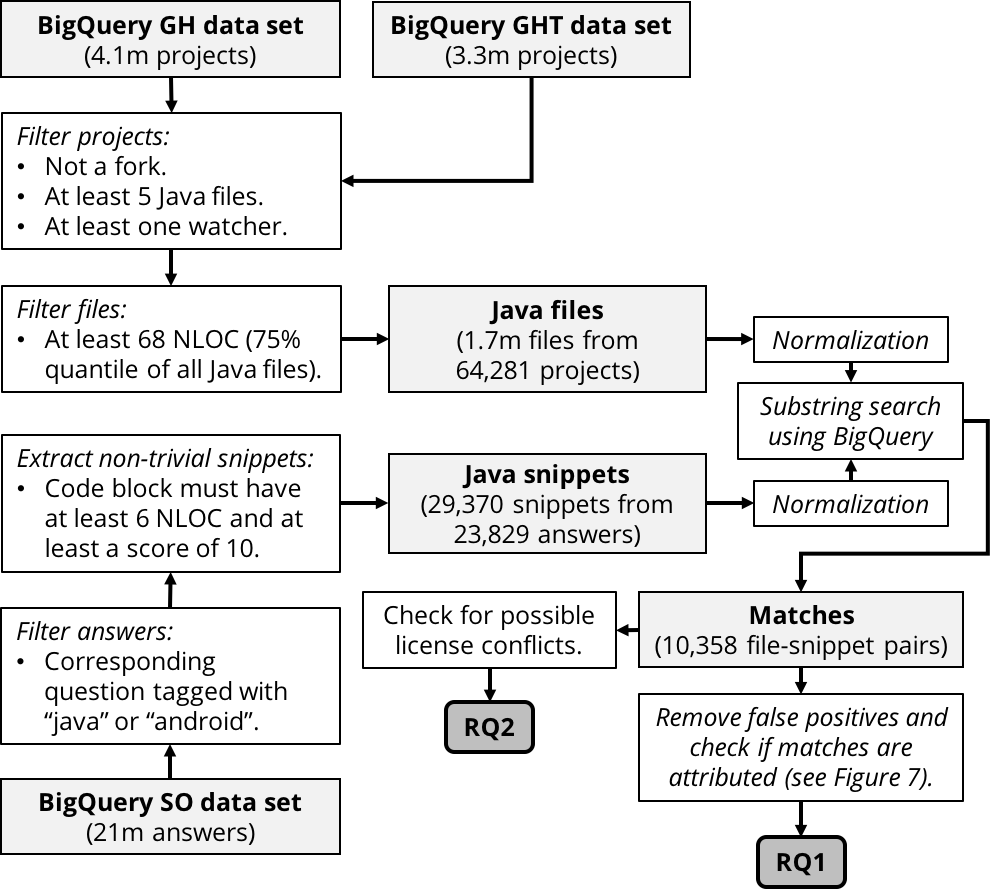} % left bottom right top
\caption{RQ1 -- Phase 3: We searched for as many exact matches of Java snippets from Stack Overflow (SO) in public GitHub (GH) projects as feasible. We filtered the GH Java projects to exclude small `toy' projects and further excluded short and unpopular SO snippets. NLOC means that we normalized the source code before we determined its length. In the end, we searched for exact matches of 29,370 snippets in 1,7m Java files (50.5 billion combinations) (time span of this phase: 03/2017--04/2017).}
\label{fig:phase3}
\end{figure}

\begin{figure}
\centering
\includegraphics[width=1\columnwidth,  trim=0.0in 0.2in 0.0in 0.2in]{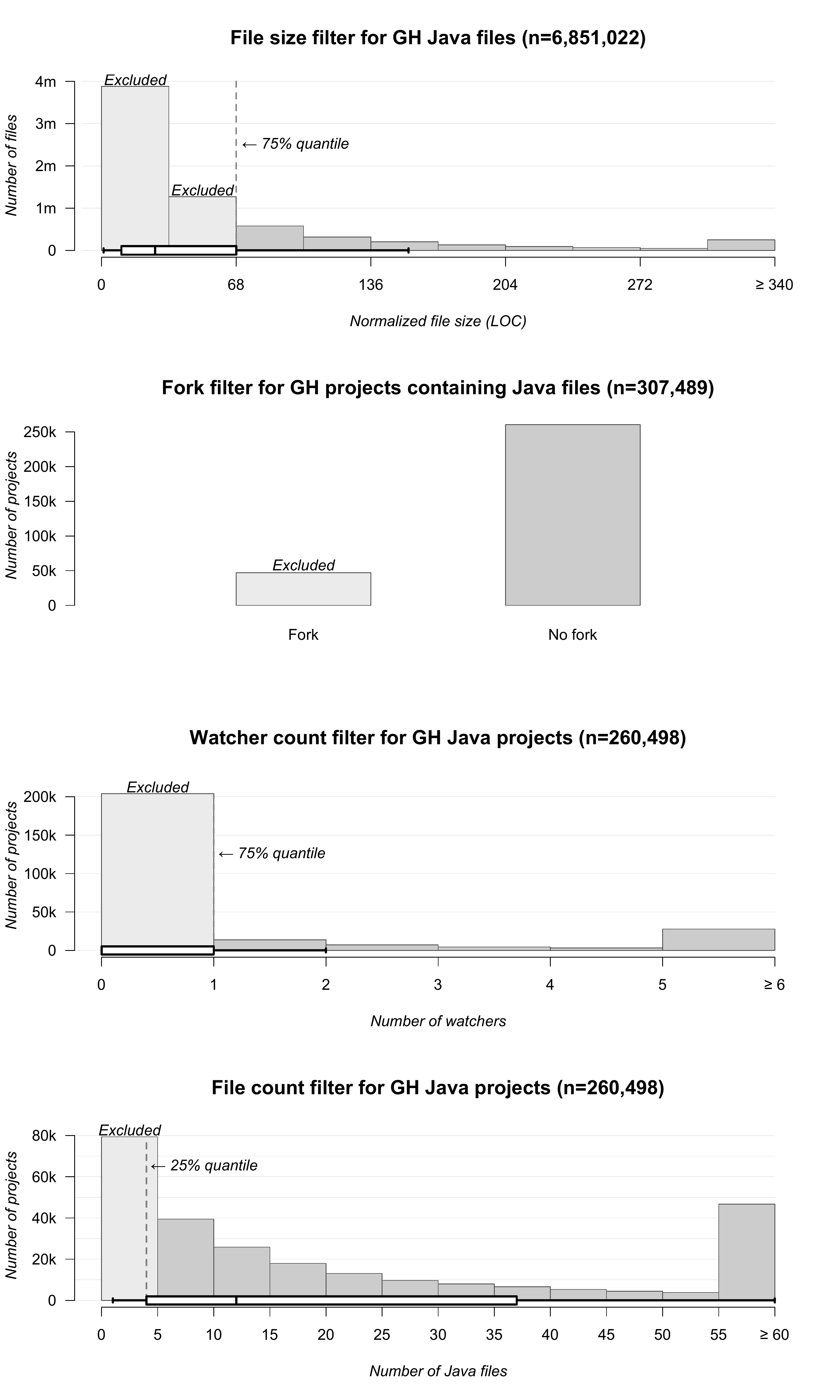} % left bottom right top
\caption{RQ1 -- Phase 3: Barplot and histograms with boxplots visualizing the applied filters to reduce the number of GitHub (GH) Java files we searched for exact matches of Stack Overflow (SO) snippets; 65 LOC was 75\% quantile of all Java files on GH; 1 watcher was 75\% quantile of all GH projects containing Java files; 4 files was 25\% quantile of all GH projects containing Java files; based on the GHTorrent BigQuery data set 2017-01-19.}
\label{fig:phase3-gh-filter-histograms}
\end{figure}

\begin{figure}
\centering
\includegraphics[width=1\columnwidth,  trim=0.0in 0.2in 0.0in 0.2in]{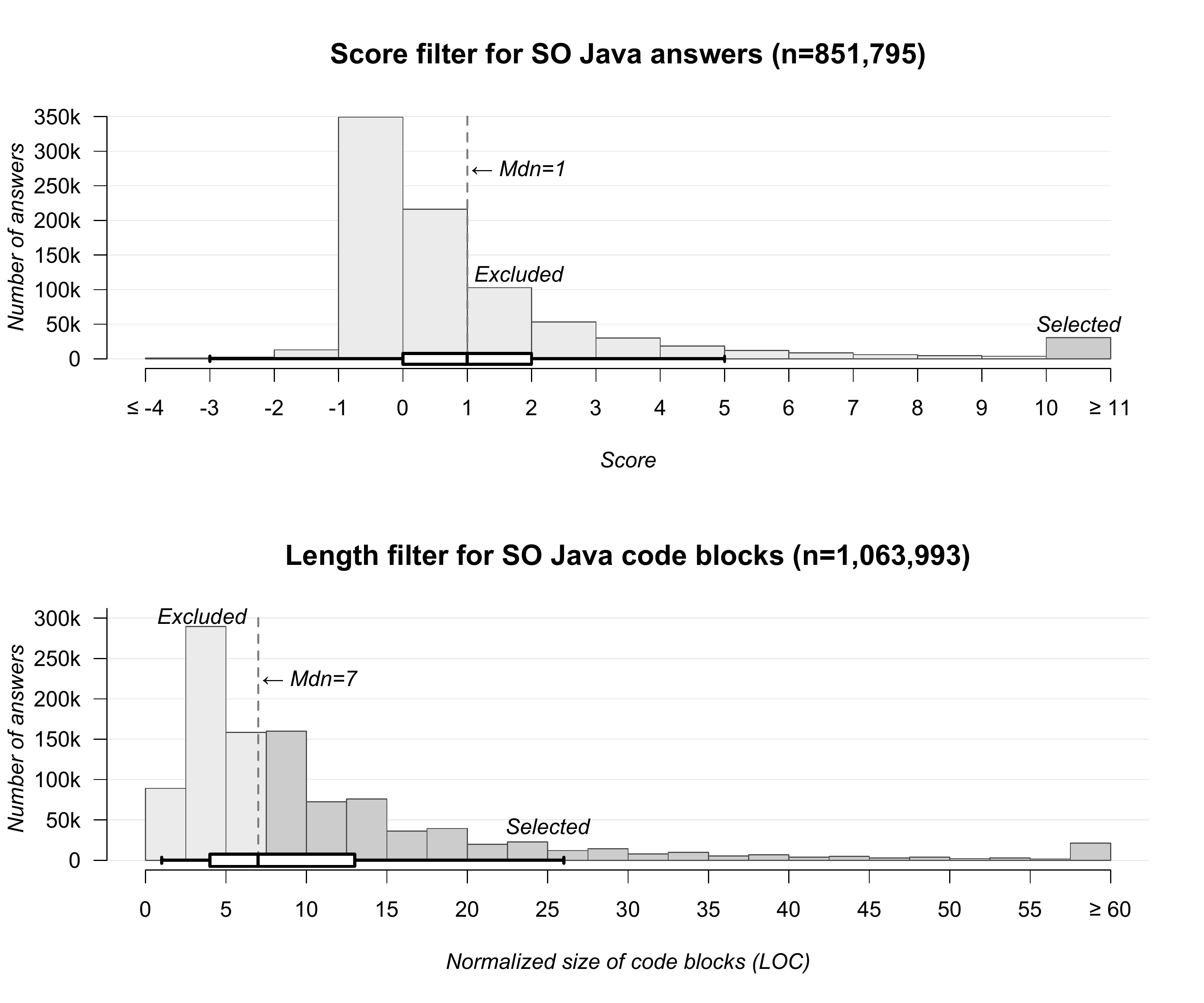} % left bottom right top
\caption{RQ1 -- Phase 3: Histograms with boxplots visualizing the applied filters to reduce the number of Java code snippets from Stack Overflow (SO) in our search for exact matches of these snippets in Java files hosted on GitHub (GH); based on the Stack Overflow BigQuery data set 2017-03-27.}
\label{fig:phase3-so-filter-histograms}
\end{figure}

Our third and last approach to answer RQ1 addressed the main shortcoming of the previous phases, which was the relatively small number of SO code snippets being analyzed.
Since the approaches of the first two phases do not scale due to the manual creation of regular expressions (phase 1) or the performance of the code clone detector (phase 2), we focused on exact matches of SO snippets in the third phase, which are easier to find.
We searched for exact matches of SO snippets in GH projects using the public BigQuery \emph{GH}, \emph{GHTorrent}, and \emph{SO} data sets~\citep{GoogleCloudPlatform2017b, GoogleCloudPlatform2017, Gousios2017} and iteratively filtered the resulting matches to exclude false positives and snippets that were also available in other sources than SO.

\subsection{Method:}

For various reasons, it is not feasible and sensible to search for all code snippets on SO in all projects on GH.
BigQuery's \textit{GH data set} consists of (almost) all public files on GitHub, which includes many small software projects of single users and also repositories that are not used for hosting software projects~\citep{KalliamvakouGousiosOthers2014, MunaiahKrohOthers2017}.
Moreover, very small code snippets from SO would produce many false positives and it is likely that such snippets are not protected by copyright.
Since there is no ``international standard for originality''~\citep{CreativeCommonsCorporation2017b} that defines when a code snippet is protected by copyright, we based our filter on the length distribution of SO code snippets and only selected snippets having a certain size.
We thus used the length of the snippets as a proxy variable for their originality. 

Another reason to filter the snippets and projects was to reduce the complexity to make the search for exact matches feasible.
For every filter we applied, we considered the distribution of values for the corresponding variables.
Figure~\ref{fig:phase3} visualizes how we filtered the Java files and the Java snippets to reduce the number of combinations to a level that allowed us to employ BigQuery's \texttt{STRPOS} function to search for matches of the snippets in the files.

We first used the BigQuery \emph{GHTorrent data set} to filter out repositories that were forks of other repositories (see Figure~\ref{fig:phase3-gh-filter-histograms}).
Then, we excluded projects with less than five Java files and less than one watcher to get rid of the many `toy' projects hosted on GitHub~\citep{KalliamvakouGousiosOthers2014}.
Afterwards, we normalized the contents of the remaining Java files by removing all lines with import or package statements, deleting comments, and normalizing the whitespaces (removing empty lines and converting multiple newline characters to one newline character).
We excluded Java files having less than 68 normalized lines of code, which was the 75\% quantile for all Java files, resulting in a sample of 1,720,587 files from 64,281 projects.
To improve the substring matching, we then further normalized the file contents by converting the characters to lower case and deleting semicolons, curly and regular braces, and all whitespace characters.

\begin{figure}
\centering
\includegraphics[width=0.98\textwidth,  trim=0.0in 0.0in 0.0in 0.0in]{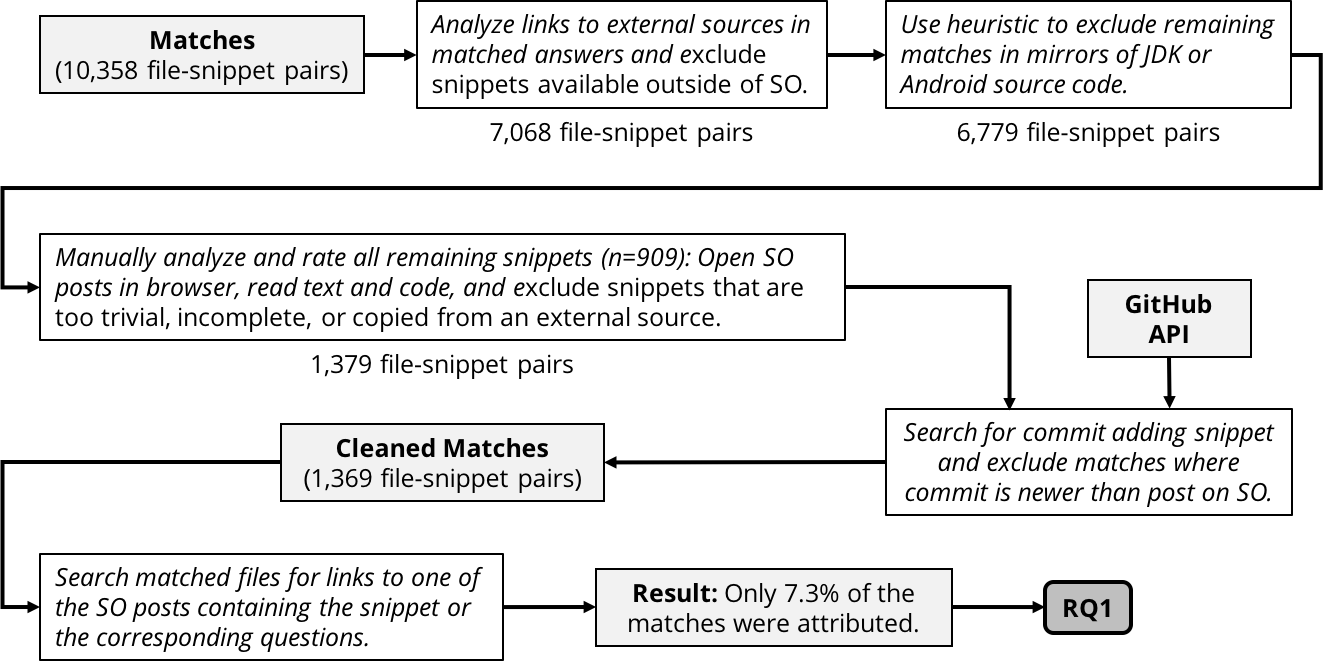}
\caption{RQ1 -- Phase 3: Our workflow to remove false positive matches and snippets available in other sources than the SO post.}
\label{fig:phase3-cleanup}
\end{figure}

To retrieve the Java snippets for the substring search, we first extracted all answers to questions tagged with \texttt{java} or \texttt{android} from the BigQuery \emph{SO data set}.
Then, we analyzed the score of the answers.
% and observed that 68\% of the answers had a score of less than two.
To concentrate on answers that gained a certain degree of attention, we excluded all answers with a score of less than ten (see Figure~\ref{fig:phase3-so-filter-histograms}).
Then, we used the Markdown representation of the posts to identify and extract continuous code blocks from the answers.
We normalized the snippets exactly like the Java files and analyzed their length.
As we were interested in non-trivial code snippets, we removed all code blocks with less than 6 normalized lines, which excluded about a third of the code blocks.
We then further normalized the snippets, analogously to the file contents.
To illustrate the normalization, we provide the normalized version of the snippet presented in Section~\ref{sec:phase1-results}:

\vspace{\baselineskip}

\begin{java}
stringhumanreadablebytecountlongbytes,booleansiintunit=si?1000:
1024ifbytes<unitreturnbytes+"b"intexp=intmath.logbytes/math.log
unitstringpre=si?"kmgtpe":"kmgtpe".charatexp-1+si?"":"i"return
string.format"%.1f%sb",bytes/math.pow(unit,exp),pre
\end{java}

After the substring search was complete, we employed different approaches to exclude false positives from the matches (see Figure~\ref{fig:phase3-cleanup}):
First, we manually investigated all matches of SO answers with links to external sources and checked whether the code on GH may have also been copied from there.
We observed that many repositories contained mirrors of the \emph{OpenJDK} or the \emph{Android} source code. To exclude matches involving files from those sources, we used a heuristic based on path names.
We then manually investigated the SO posts of all remaining snippets and excluded snippets that we either rated as being too trivial or incomplete, or where the post indicated that the snippet has been copied from a third source (without providing a link).
As motivated above, there is no international standard defining when a snippet is original enough to be copyrightable.
Our notion of `too trivial' included snippets that consist only of a few API calls or that are so simple that another developer would likely come up with the same code.
Moreover, we checked if the snippets are `complete' in the way that they are ready to copy-and-paste, without substantial modifications.
Since those judgments are, to some degree, subjective, we tried to mitigate a possible bias by discussing borderline cases.

%\todo{successful for how many snippets?}
As a last step, we employed the \textit{GitHub API} and searched for the commit that added the snippet to the repo. We then removed matches where the commit on GH was older than the post on SO.

%We observed that some non-trivial code blocks were present in different answers,  

\subsection{Results:}

After removing potential false positive matches and snippets that were also available in other sources, the result set consisted of 1,369 snippet-file pairs.
For the remaining matches, we are quite confident that they are in fact clones of the SO snippets and are not copied from a different source. 
Only 104 (7.6\%) of the snippet-file pairs were attributed using a link to one of the SO posts containing the snippet (some snippets were present in more than one post) or to the corresponding questions.
We found exact matches of SO snippets in 764 (1.19\%) of the 64,281 analyzed GH repositories.
Using the \textit{BigQuery GH data set}, we also analyzed the licenses of those repositories.
The results of this analysis can be found in Section~\ref{sec:licensing-conflicts}.

\begin{normalbox}
\textbf{Usage Without Attribution (RQ1 -- Phase 3):}
We searched for exact matches of 23,829 Java snippets from SO in 64,281 GH projects and excluded snippets available in external sources. Only 7.6\% of the 1,369 matches were attributed.
\end{normalbox}

\section{Summary (RQ1 -- Phases 1--3)}
\label{sec:summary}

In this section, we summarize our results from all three approaches to quantify the amount of unattributed usages of non-trivial Java code snippets from SO in public GH projects (RQ1).
For each phase, Table~\ref{tab:summary} provides an overview on: (1) the number of distinct references to answers and questions in the analyzed Java files, (2) the number of Java files and repositories we analyzed, and (3) the number and ratio of attributed usages of SO code snippets in the analyzed files.

Taking all three phases into account, we consider one quarter to be a reasonable upper bound for the ratio of attributed usages of SO Java snippets in GH files (see Table~\ref{tab:summary}, column Files $\rightarrow$ \textsc{attr}).
Between 3.3\% and 11.9\% of the analyzed repositories contained references to Stack Overflow questions or answers (Repositories $\rightarrow$ \textsc{Ref}).
The table further shows the number of distinct analyzed files (Files $\rightarrow$ \textsc{count}), along with the percentage of files containing a reference to SO (Files $\rightarrow$ \textsc{ref}).

Moreover, Table~\ref{tab:summary} lists the number of distinct references to SO posts we identified in each phase (References), where distinct means that we counted copied files only once. If one file contained the same URL several times, we also counted it only once.
In our analysis, we ignored URLs that were either malformed or referred to other content on SO such as tags or users.
For instance, of all SO URLs we found in the first phase, 2.16\% did not refer to a question or an answer.

\begin{table}
\caption{Summary of results from phases 1 to 3: Distinct references to answers (A) or questions (Q) on Stack Overflow (SO) in the Java files from GitHub analyzed in each phase; number of analyzed files and repositories, files/repos containing a reference to SO, files/repos containing a copy of a SO snippet, attributed copies of SO snippets.}
\begin{tabular}{crrrrrrrrr}
\hline\noalign{\smallskip}
\multicolumn{1}{c}{\multirow{2}{*}{Ph.}} & \multicolumn{2}{c}{References} & \multicolumn{4}{c}{Files} & \multicolumn{3}{c}{Repositories} \\
& \multicolumn{1}{c}{\textsc{A}} & \multicolumn{1}{c}{\textsc{Q}} & \multicolumn{1}{c}{\textsc{Count}} & \multicolumn{1}{c}{\textsc{Ref}} & \multicolumn{1}{c}{\textsc{Copy}} & \multicolumn{1}{c}{\textsc{Attr}} & \multicolumn{1}{c}{\textsc{Count}} & \multicolumn{1}{c}{\textsc{Ref}} & \multicolumn{1}{c}{\textsc{Copy}} \\
\noalign{\smallskip}\hline\noalign{\smallskip}
\multirow{2}{*}{1} & 5,014 & 16,298 & 13.3m & 18,605 & 4,198 & 402 & 336k & 11,086 & 3,291 \\ % 13.3m -> 13,311,301; 336k -> 336,028
& 23.5\% & 76.5\% & & 0.09\% & 0.03\% & \graybg9.6\% & & 3.3\% & 1.0\% \\
\noalign{\smallskip}\hline
\multirow{2}{*}{2} & 209 & 463 & 445k & 634 & 297 & 70 & 2,313 & 274 & 199\\
& 31.1\% & 68.9\% & & 0.14\% & 0.07\% & \graybg23.6\% & & 11.9\% & 8.6\% \\ % 445k -> 444,611
\noalign{\smallskip}\hline
\multirow{2}{*}{3} & 1,551 &  4,843 & 1.7m & 5,354 & 1,369 & 104 & 64,281 & 3,536 & 1,332 \\ % 1.7m -> 1,720,587
& 24.3\% & 75.7\% & & 0.31\% & 0.08\% & \graybg7.6\% & & 5.5\% & 2.1\% \\
\noalign{\smallskip}\hline
\end{tabular}
\label{tab:summary}
\end{table}

Generally, developers were more likely to refer to a question, that is to the whole thread, compared to a particular answer.
%For question links, we do not know which answer, and hence which snippet, the developer exactly referred to.
%For our analysis, we focused on answers, because in case the question is referenced, we do not know which snippet (if any) has been copied.
In the first phase, only 0.09\% of the analyzed files and only 3.3\% of the analyzed projects contained a reference to SO.
However, these results include all public files on GH in the \textit{BigQuery data set}, which includes many small software projects of single users and also repositories that are not used for hosting software projects~\citep{KalliamvakouGousiosOthers2014, MunaiahKrohOthers2017}.

%\todo{add number of SO users and SO site visits over time to figures}
%\todo{How many answers contained code snippets? Size (characters/LOC)?}

%%% TODO %%%
%\section{Age of Stack Overflow content when used in GitHub projects}	 	
%%% TODO %%%

\section{Frequency of Licensing Conflicts (RQ2)}
\label{sec:licensing-conflicts}

To assess how often the license of repositories containing code copied from SO conflicts with SO's license (RQ2), we retrieved the license of all repositories previously identified as containing code from SO.
To this end, we employed the \textit{GitHub API} (phase 1+2) and the \textit{BigQuery GH data set} (phase 3).
Tables~\ref{tab:repos-license-phase1}, \ref{tab:repos-license-phase2}, and \ref{tab:repos-license-phase3} show the five most common licenses of the matched repositories from each phase.
We provide the complete lists as supplementary material~\citep{Baltes2018e}.
Between 1.82\% (attributed matches in phase 1) and 38.9\% (unattributed matches in phase 2) of the matched repositories did not provide a license (or at least none that the GitHub API was able to identify).
The relatively large number of repositories without a license may seem unusual, but it is in line with a recent study by Meloca et al., who found that it is common in open source projects to not provide a license~\citep{MelocaPintoOthers2018}.
Moreover, some files or directories could have their own license, differing from the repository's license. 
As we never observed this for the files we manually analyzed, we relied on the repository's license for our analysis. 

None of the analyzed projects used the CC BY-SA 3.0 or the CC BY-SA 4.0 license, which would be share-alike compatible with the content from SO.
One could leverage the upwards compatibility of CC BY-SA 3.0 and CC BY-SA 4.0~\citep{CreativeCommonsCorporation2017} and the share-alike compatibility of CC BY-SA 4.0 and GPL 3.0 to achieve a share-alike compatibility of CC BY-SA 3.0 and GPL 3.0. %, but it is unclear if this transitive use of the share-alike compatibility is legitimate.
Still, only 60 (1.8\% of all matched repos in phase 1), 6 (3.0\% of all matched repos in phase 2), respectively 19 (1.4\% of all matched repos in phase 3) repositories were licensed under GPL 3.0 and attributed the code copied from SO as required by the license.
Thus, only those 85 repositories (1.8\% of all matched repos) may have used the snippets in a way compatible with CC BY-SA 3.0, meaning with attribution and with a share-alike compatible license.

\begin{normalbox}
\textbf{Frequency of Licensing Conflicts (RQ2):}
At most 1.8\% of all analyzed repositories containing code from SO used the code in a way compatible with CC BY-SA 3.0.
\end{normalbox}

\begin{table}
\caption{Five most common licenses of GitHub repositories matched in phase 1 containing attributed or unattributed copies of code snippets from Stack Overflow.}
\begin{tabular}{lrr}
\hline\noalign{\smallskip}
\multicolumn{1}{c}{\multirow{2}{*}{SPDX license name}} & \multicolumn{2}{c}{Number of repos containing a SO code snippet clone that was:} \\
 & \multicolumn{1}{c}{unattributed ($n=2,962$)} & \multicolumn{1}{c}{attributed ($n=329$)} \\
\noalign{\smallskip}\hline\noalign{\smallskip}
Apache-2.0 & 921 (31.1\%) & 99 (30.1\%) \\
MIT & 621 (21.0\%) & 72 (21.9\%) \\
GPL-3.0 & 435 (14.7\%) & \graybg60 (18.2\%) \\
GPL-2.0 & 284 (9.6\%) & 21 (6.4\%) \\
BSD-3-Clause & 82 (2.8\%) & 9 (2.7\%) \\
\noalign{\smallskip}\hline
\end{tabular}
\label{tab:repos-license-phase1}
\end{table}

\begin{table}
\caption{Five most common licenses of GitHub repositories matched in phase 2 containing attributed or unattributed copies of code snippets from Stack Overflow.}
\begin{tabular}{lrr}
\hline\noalign{\smallskip}
\multicolumn{1}{c}{\multirow{2}{*}{SPDX license name}} & \multicolumn{2}{c}{Number of repos containing a SO code snippet clone that was:} \\
 & \multicolumn{1}{c}{unattributed ($n=144$)} & \multicolumn{1}{c}{attributed ($n=55$)} \\
\noalign{\smallskip}\hline\noalign{\smallskip}
None & 56 (38.9\%) & 18 (32.7\%) \\
Apache-2.0 & 33 (22.9\%) & 15 (27.3\%) \\
GPL-3.0 & 17 (11.8\%) & \graybg6 (10.9\%) \\
MIT & 6 \phantom{0}(4.2\%) & 4 \phantom{0}(7.3\%) \\
GPL-2.0 & 4 \phantom{0}(2.8\%) & 2 \phantom{0}(3.6\%) \\
\noalign{\smallskip}\hline
\end{tabular}
\label{tab:repos-license-phase2}
\end{table}

\begin{table}
\caption{Five most common licenses of GitHub repositories matched in phase 3 containing attributed or unattributed copies of code snippets from Stack Overflow.}
\begin{tabular}{lrr}
\hline\noalign{\smallskip}
\multicolumn{1}{c}{\multirow{2}{*}{SPDX license name}} & \multicolumn{2}{c}{Number of repos containing a SO code snippet clone that was:} \\
 & \multicolumn{1}{c}{unattributed ($n=1,169$)} & \multicolumn{1}{c}{attributed ($n=163$)} \\
\noalign{\smallskip}\hline\noalign{\smallskip}
Apache-2.0 & 353 (30.2\%) & 36 (37.4\%) \\
MIT & 239 (20.4\%) & 25 (15.3\%) \\
GPL-3.0 & 211 (18.0\%) & \graybg19 (11.7\%) \\
None & 153 (13.1\%) & 61 (37.4\%) \\
GPL-2.0 & 89 (7.61\%) & 8\phantom{0} (4.9\%) \\
\noalign{\smallskip}\hline
\end{tabular}
\label{tab:repos-license-phase3}
\end{table}

\section{Adherence to Attribution Requirements (RQ3)}
\label{sec:attribution-requirements}

Until May 2018, SO defined certain attribution requirements in their terms of service~\citep{StackExchangeInc2018b}. The following information was required when content from SO was republished:
\begin{enumerate}
	\item A visual indication that the content is from SO,
	\item a hyperlink directly to the original question,
	\item the authors' names for every question and answer,
	\item a hyperlink for each author to their profile page on SO.
\end{enumerate}
% https://wiki.creativecommons.org/wiki/Best_practices_for_attribution
% CC FAQ: "Can I insist on the exact placement of the attribution credit? No. CC licenses allow for flexibility in the way credit is provided depending on the medium, means, and context in which a licensee is redistributing licensed material. For example, providing attribution to the creator when using licensed material in a blog post may be different than doing so in a video remix. This flexibility facilitates compliance by licensees and reduces uncertainty about different types of reuse—minimizing the risk that overly onerous and inflexible attribution requirements are simply disregarded."
% ...
% "We also advise against modifying our licenses through indirect means, such as in your terms of service. A modified license very likely will not be compatible with the same CC license (unmodified) applied to other material. This would prevent licensees from using, combining, or remixing content under your customized license with other content under the same or compatible CC licenses."
%The above requirements seem to mainly target republication of whole threads, but as they apply for all content on SO, they also apply for code snippets.
However, Creative Commons states that one cannot ``insist on the exact placement of the attribution credit''~\citep{CreativeCommonsCorporation2017b}.
Thus, it is unclear if the above attribution requirements can actually be enforced by SO.
Moreover, Creative Commons points to the fact that altering a CC license through ``indirect means'', like terms of service, could make the modified license incompatible with the CC license itself.
Nevertheless, our goal was to find out to what degree developers adhere to SO's attribution requirements when they refer to SO posts in source code comments (RQ3).
As described in the introduction, SO's revised terms of service do not mention the attribution requirements anymore, but they are still linked from the footer of the website (visible for each thread) and from the help page.
Regardless of the enforceability of those requirements, the following analysis provides valuable insights into how GH users reference code copied or adapted from SO answers.  

\subsection{Method:}
In the first phase of our research, we identified 2,443 distinct SO answers that were referenced from at least one Java file on GH.
We drew a random sample of those answers to investigate how GH users attribute code snippets from SO ($n=100$).
If a URL in the sample had multiple references, we randomly chose one of them.

To determine the \emph{margin of error} for this sample, we first calculated the \emph{standard error} (SE)~\citep{Agresti2007}, assuming that the probability to observe a correct attribution is 50\% ($p = 0.5$).
For our sample size of 100, this probability yields a standard error of $0.05$.
In fact, we did not observe a correct attribution in any case (see Section~\ref{sec:attribution-requirements-results}), thus the actual probability is likely to be much lower.
%\begin{math}SE = \sqrt{\frac{p(1-p)}{n}} = \sqrt{\frac{0.1 \cdot 0.9}{100}} = 0.03\end{math}
Based on the standard error and a confidence level of 95\% ($\alpha = 0.05$), we calculated the margin of error by multiplying the \emph{z-score}~\citep{Agresti2007, BartlettKotrlikOthers2001, Cochran1977}: $z(\nicefrac{\alpha}{2})\cdot SE = 0.10$.
% Agresti2007: z(a/2) denotes the standard normal percentile having right-tail probability equal to a/2
% Wikipedia: The margin of error is usually defined as the "radius" (or half the width) of a confidence interval for a particular statistic from a survey.
Thus, with the above-mentioned assumptions, the margin of error for our estimation of references not adhering to the attribution requirements is 10 percentage points.
This means that, even if we did not observe a correct attribution in any of the sampled cases, there could still be up to 10\% references adhering to the attribution requirements (confidence level 95\%).

We manually extracted the snippets from SO and the referencing code from GH and coded how and where the user attributed the snippet and if he or she just copied, or also adapted, the snippet.
We provide the extracted snippets, files, and our coding as supplementary material~\citep{Baltes2018e}.

\subsection{Results:}
\label{sec:attribution-requirements-results}

Of the 100 referenced answers we analyzed, 12 were conceptual and contained no code suitable for copying and pasting.
Three references did not exist anymore when we tried to access the files (file or repository moved or deleted).
Most references (89) included only the URL to the answer in a comment, eight references further included the username of the author, e.g.:
\begin{java}
/**
  * Converts a double to a String in [...]
  * Based on Stack Overflow answer by corsiKa at http://Stack Overflow.com/a/5036540 [...]
**/
\end{java}

To introduce their reference, most developers (62) used formulations like `code from', `based on', or `adapted from'; 35 users only added the SO URL without any further comment.
For the majority of references (60), the code had been adapted (e.g., variables renamed).
In two of those cases, the comment named an additional source for the copied code beside the SO answer.
In about a quarter of cases (22), the code had been copied without any modifications.
In two references, the SO answer was only included to show an alternative solution to a problem.
Further, one GH user included a link to advertise his or her own answer on SO.

About half of the references were made in regular source code comments, most of which were placed above the copied snippet (only two were inline comments behind a statement); 41 references were JavaDoc comments for classes, methods, or class variables. 
It is unclear what SO considers a proper ``visual indication'' that the content is from SO (required according to the terms of service).
Still, only 11 references explicitly mentioned the term `Stack Overflow' (or other spellings like `StackOverflow' or `S.O.') in their comment.
Further, none of the comments included a link to the author's profile page, which was also required according to SO's terms of service.

\begin{normalbox}
\textbf{Adherence to Attribution Requirements (RQ3):}
Most comments referencing code snippets copied or adapted from Stack Overflow included only a link to the corresponding answer without naming the author of the code.
%In a quarter of the analyzed references, the snippet was copied without modifications, but in most cases the code had been changed.
No comment included a link to the author's profile page and only 11 out of 97 analyzed comments explicitly named SO as the source.
In summary, none of the analyzed references fulfilled the four attribution requirements defined by SO.
\end{normalbox}

\section{Developers' Awareness Regarding SO's Licensing (RQ4)}
\label{sec:awareness-survey}

%\todo{name matches that were removed before contacting participants}

To complement our estimation of unattributed usages of SO code snippets in GH projects, we conducted a second online survey investigating the awareness of GH developers regarding the licensing of SO content.
We further used this survey to reveal false positives in our analysis. % for RQ1.
Moreover, we contacted the authors of the ten most frequently referenced SO Java answers, identified in phase 1 (see Section~\ref{sec:phase1}), and asked them about their view on the snippets' licensing situation.

\subsection{Method:}

For the online survey, we derived a sampling frame from the GH Java repositories that contained at least one file with a clone of the ten most frequently referenced SO Java snippets identified in the first phase of our research (see Section~\ref{sec:phase1}).
We retrieved the owners for those repositories using our \textit{api-retriever} tool~\citep{Baltes2017c}, which utilizes the \textit{GitHub API}.
We then filtered the GH users and organizations to only include the ones having a public email address on their profile page.
Of the 3,031 email addresses we collected, 2,165 were valid.
In a first iteration, we contacted all 211 organizations with valid email addresses and received 20 answers (9.5\% response rate).
%Based on the feedback from the first iteration we added one more question about the ShareAlike requirement of the SO license.
%Because we initially derived the sampling frame directly from the BigQuery GitHub data set that contained some (but not all) forked repositories~\citep{Hoffa16}, we utilized the GitHub API to excluded owners of forked repositories.
For the second iteration, we removed owners of forked repositories and then contacted 528 developers, receiving 67 responses (12.7\% response rate).
In both iterations, we informed participants about all matches we found in their repositories and asked them for one randomly selected match if the code has actually been copied from SO.
We provide the questionnaire, the analysis scripts, as well as all closed-ended responses as supplementary material~\citep{Baltes2018e}.

To contact the authors of the ten most frequently referenced SO Java answers, we checked their SO profile and searched for their user name on the web.
We collected the email address of two authors from their personal website and of five authors from their GH profile, but we were not able to retrieve the email address of four authors.
Please note that we have eleven authors in total, because the answer ranked fifth actually pointed to a question and we selected two answers for that question (see Table~\ref{tab:java-top10-1}).
The email we sent to those authors contained three questions: One asking about their awareness regarding SO's licensing, one asking about an additional source for the snippet, and one asking whether they care about attribution for the particular snippet.

\subsection{Results:}

In total, 87 users responded to the online survey (11.8\% response rate). % (67+20)/(211+528)*100
Beside the survey responses, we received many emails from participants, thanking us that we informed them about the licensing of SO code snippets and in particular about unattributed usages in their projects.
One participant, for instance, wrote that his/her team replaced the matched snippet in the repo due to the \textit{share-alike} requirement of SO's license, which they ``ignored until [we] called [their] attention.''
Another participant informed us that the match was in a mirror of the OpenJDK 9 Mercurial repo that was part of the GH repo we analyzed.
We informed the OpenJDK team and they replaced the code due to legal concerns.
In the corresponding bug description, the author points to possible legal issues and the fact that it is ``not a good practice'' to copy code from SO~\citep{Fazunenko2016}.

Similar to our preliminary study (see Section~\ref{sec:preliminary-study}), the majority of respondents (62\%) reported their main software development role to be \textit{software developer}, but there were fewer \textit{software architects} (8\%).
The average age of the participants who reported their age (n=65) was 30.3 years ($SD{=}9.4$) and they had an average programming experience of 11.7 years ($SD{=}8.9$).
% Origin (Europe, North America, etc.)
Again, most users answered that they use SO (80\%) and GH (61\%) for both private and work-related projects; almost one third of them use GH only for private projects (28\%).

As mentioned above, we asked participants for one match that we found in their repo whether the code has actually been copied from SO.
Of the 74 participants who answered to that question, 43 answered `Yes' (58\%), 20 answered `No' (27\%), and 11 (15\%) answered `I don't know'.
Of the 20 participants who answered that the snippet has not been copied from SO, seven claimed they wrote the code themselves, two claimed that a team member wrote it, and 11 answered that they copied it from another source.
We manually inspected those matches:
Eight of them (10.8\%) were indeed relatively short and thus likely to be false positives.
To us, the other 12 matches looked like copies of the SO snippets.
Three of them were copies of a SO snippet that was itself a copy of another SO snippet; five matches were also available in external sources like personal blogs (one licensed under the Apache License 2.0, the others were not licensed).
Some of the participants who answered that they wrote the code themselves may either not remember copying the code or their answer could be affected by a social desirability bias~\citep{Nederhof1985}.
To mitigate the former and to enable tracing the source of code copied from SO, developers should add a comment with a link to SO as motivated in the introduction.

%We made another interesting observation while inspecting the matches: 
%Three matches were copies of snippet \texttt{a/9655181}, which is again a copy of snippet \texttt{a/332079}.

%Three matches were copies of a SO snippet that was itself a copy of another SO snippet.
%%One match, \texttt{a/9655181}, is a revised version of an old blog post from 2004, which was licensed under the Apache License 2.0.
%One match was a revised version of an old blog post from 2004, licensed under the Apache License 2.0.
%%Moreover, two matches were copies of \texttt{a/3758880}, which is a copy of a post on a personal website of the SO author.
%Moreover, two matches were copies of a snippet that the SO author copied from his personal website (no license provided).
%%For this blog post, the author holds the copyright (no license provided).
%All these snippets can still be used under CC BY-SA 3.0 as well. %~\citep{JonasCz16}, but may additionally licensed under another license. % if the code as been copied or if the author adds a copyright remark to his or her post.

We asked the participants if they knew that SO's license requires them to attribute code copied from posts and in particular if they knew that content on SO is licensed under CC BY-SA 3.0.
Regarding the need to attribute content copied from SO, 28 participants (32\%) were aware of it, 58 (67\%) not, and one preferred not to answer.
As to the specific license, the answers were similar: 21 participants (24\%) were aware of it, 65 (75\%) not, and one preferred not to answer.
The attribution requirements from SO's terms of service were even more unfamiliar to the participants: 11 (13\%) knew them, 73 (84\%) not, and 3 preferred not to answer.
Thus, we can conclude that most developers are not aware of the licensing of code published on SO and the implications of this licensing.

With regard to the attribution practice, we asked the same questions as in the preliminary study (see Section~\ref{sec:preliminary-study}) and got similar results:
Again, not attributing the code when coping from Stack Overflow was a common practice (41\%).
This time, we asked if respondents referred to the question or a specific answer on SO in case they added a source code comment.
Twelve participants preferred not to answer this question, seven named other information they included in the comment.
Unlike the results from our quantitative analysis of attributed usages would suggest (see Section~\ref{sec:summary}), participants more frequently reported that they referred to an answer (30\%) than to a question (13\%).
One reason for this could be that many of the references to questions refer to conceptual threads on SO that do not contain code suitable for copying and pasting.

Of the seven contacted SO authors, four answered.
Three were not aware of SO's licensing when they posted their answer, one was ``vaguely aware''.
All respondents indicated that they do not know any other source for the code in their answers (except for the ones listed in Table~\ref{tab:java-top10-1}).
One author answered: ``I invented it [the snippet] there and then. I would assume any other source would be a copy from SO.''
A different author wrote that his answer was ``informed by, but not copied directly from, other Stack Overflow posts''.
Three authors responded that they do not care about attribution for this particular content and one author answered that he does ``not really'' care.
The same author further noted that ``it's Stack Overflow that collects the money for the ads. HOWEVER, if the situation would have been the same for an article on [URL removed] which I run myself, I would care deeply about attribution.''
Another author answered that he does not have the ``desire to 'own' the information, only to share it''.
Those two comments, together with the discussion around SO's attempt to change the license for code snippets (see Section~\ref{sec:legal-situation}), show that developers have diverse opinions about the attribution requirement.
Further investigating the reasons to (not) care about attribution of online code snippets is an interesting direction for future work.
%``If the answer contains a substantial original contribution of ready-to-use code, then attribution should be required by default.''~\citep{StackExchangeMeta2015b}.
%``Change 1 was made to accommodate contributors who want credit, plus to help developers identify the provenance of a Stack Overflow code snippet when they find it integrated into a project.''~\citep{StackExchangeMeta2016}

\begin{normalbox}
\textbf{Awareness of Licensing (RQ4):}
Most developers answering to the online survey were not aware of the licensing of code published on SO and its implications.
75\% of the participants did not know that content on SO is licensed under CC BY-SA 3.0 and 67\% did not know that attribution is required.
%, and 84\% were not aware of the attribution requirements defined in the SO terms of service.
Not attributing the code when coping code from SO was a common practice (41\%).
\end{normalbox}

\section{Limitations and Verifiability}
\label{sec:limitations}

%\todo{dual licensing, Even if users add a copyright remark to their posts or publish the code elsewhere under a different license, it can still be used under CC BY-SA 3.0 as well~\citep{SOM16}.}
%\todo{erwähnen, dass attribution requirements from ToS may not be valid according to CC}
%\todo{Reviewer: Not accounted for cases where code on GH is older than code on SO.}
%\todo{Clones between snippets (e.g., so-answer-140861 in so-answer-25462286)}

The main limitation of our research is the focus on Java, because the attribution practice may differ between programming languages.
%As described in Section~\ref{sec:quantitative-analysis-I+III}, the attribution practice differs between the analyzed programming languages.
Thus, the generalizability of our results to other programming languages is limited.
To answer RQ1, we used three different approaches, optimized for precision and always chose conservative estimates.
Thus, we do not see the construct validity of our research to be impaired.
%One limitation of the configuration we used for PMD is that we were only able to detect type-1 clones; the ``dark figure'' of copies is likely to be higher than what we found.
%Thus, the ``dark figure'' of copies is likely to be higher than what we found.
For the first two phases, we only considered a relatively small sample of snippets compared to all available snippets on SO, but we still found a considerable number of files with copies. 
The number of attributions was even smaller in the third phase, where we included more snippets and only searched for exact matches.
%Some regular expressions we used for finding copies were rather short (snippets 2, 5, and 9) and thus some of the matches may not necessarily be copied from SO.
%However, we manually checked the results and found no matches that looked like false positives.

Another threat to validity is that both the SO snippet as well as the matched code on GH could have a different origin.
To mitigate this threat, we analyzed and described all external sources that were linked in the SO answers.
In most cases, those sources did not provide a license, thus CC BY-SA 3.0 is the only license which applies.
Another possible issue is that if users include a license statement in their snippets on SO, they may allow a more permissive usage without attribution.
However, this was only the case in very few of the snippets we manually investigated.
%Two of the analyzed snippets contained a license statement.
%One user just repeated SO's license (CC BY-SA), the other published his code under the Apache License 2.0, which still requires a ``prominent notice'' of own changes and the origin of the code. 

\begin{figure}
\centering
\includegraphics[width=1\columnwidth,  trim=0.3in 0.6in 0.2in 0.0in]{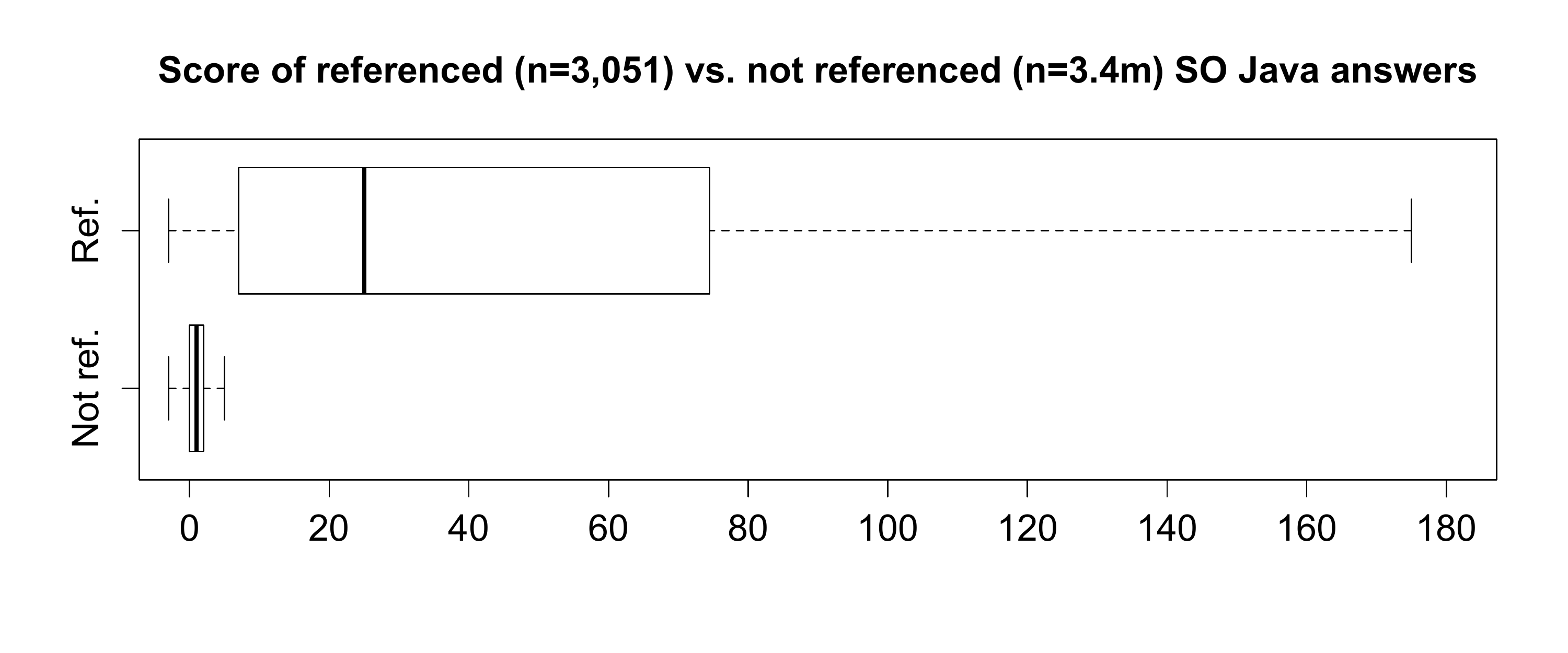} % left bottom right top
\caption{Scores of Stack Overflow (SO) Java answers referenced in public GitHub (GH) projects compared to scores of Java answers not referenced in GH projects; outliers not depicted; data retrieved from BigQuery GH and SO data sets (11/2017).}
\label{fig:score-boxplot}
\end{figure}

In phase 3 (Section~\ref{sec:phase3}), we used the length of the SO snippets as a proxy variable for their originality.
However, as mentioned above, there is no ``international standard for originality''~\citep{CreativeCommonsCorporation2017b} that defines when a code snippet is protected by copyright.
Thus, even with the threshold we chose, some of the snippets may not be copyrightable.
The survey (Section~\ref{sec:awareness-survey}) revealed that 10.8\% of the matches in phase 1 were false positives due to their short length.
We addressed this issue in phase 3 with a higher threshold for the minimum snippet length.

In phases 2 and 3, we focused on rather popular GitHub repositories to reduce complexity and exclude projects that are not ``engineered software projects''~\citep{KalliamvakouGousiosOthers2014, MunaiahKrohOthers2017}.
This approach has a very high precision, but also a relatively low recall~\citep{MunaiahKrohOthers2017}.
Thus, the results of those two phases may only be generalizable to popular projects.
Nevertheless, in popular projects the impact of licensing violations is much larger then in small personal projects.

In all phases, we focussed on rather popular SO answers. % top10 and top100
Thus, our results may not be generalizable to less popular SO answers.
Our assumption was that code from unpopular SO answers is less likely to be used in GH projects.
To assess this assumption, we utilized the \textit{BigQuery GH} and \textit{SO data sets} to compare the score of SO Java answers referenced in public GH projects to the score of Java answers not referenced on GH (see Figure~\ref{fig:score-boxplot}).
Referenced Java answers ($Mdn=25$, $M=95.33$, $SD=511.55$) had a significantly higher score than Java answers that were not referenced ($Mdn=1$, $M=2.48$, $SD=16.69$) (Wilcoxon rank sum test, $W = 9,705,800,000$, $\text{p-value} < 2.2 \cdot 10^{-16}$).

In phases 1 and 2, we did not check if the code on GH is older than the code on SO, which could indicate that the code has been copied from GH or from another source into the SO post.
In phase 3, however, we filtered out matches for which the commit adding the snippet was older than the post on SO, but this was only the case for 10 out of 1,379 matches (0.7\%).

To enable other researcher to verify our results, we provide all analysis scripts and data as supplementary material~\citep{Baltes2018e}. 
The supplementary material further includes instructions on how to apply the scripts to the data.

\section{Related Work}

%\todo{Generally, the availability of these ``cross references'' to the SO thread could support the code comprehension process~\citep{Mayrhauser94}} 
In the following, we summarize related work from different research areas, highlight connections to our study, and point to directions for future work.

\subsection{Stack Overflow and GitHub}

Over the past years, there have been various research papers on leveraging knowledge from SO, e.g., to support developers by automating the search~\citep{PonzanelliBacchelliOthers2013, CampbellTreude2017} or by augmenting API documentation~\citep{TreudeRobillard2016}.
Moreover, different tools have been developed to help developers finding code examples on the web~\citep{ZagalskyBarzilayOthers2012, BrandtDontchevaOthers2010}.
However, researchers rarely mentioned the complex licensing and copyright situation when building tools to support code reuse from the web, and in particular from SO.
Since our study indicates that many developers are not aware of SO's license and its implications (see Section~\ref{sec:awareness-survey}), future tools should inform developers about this aspect.

Regarding the populations of SO and GH users, studies described properties such as gender~\citep{VasilescuCapiluppiOthers2012}, age~\citep{MorrisonMurphyHill2013}, and geographic location~\citep{SchenkLungu2013}.
Wang et al.~\citep{WangLoDavidOthers2013} analyzed the asking and answering behavior of developers on SO and found that most developers only answer or ask one question and only 8\% answer more than 5 questions.
Bosu et al. analyzed how reputation is build on SO and provide recommendations for contributors~\citep{BosuCorleyOthers2013}.
Xia et al.~\citep{XiaBaoOthers2017} found that it is common for developers to search for reusable code snippets on the web, which is in line with Sojer and Henkel's results from an earlier study~\citep{SojerHenkel2011}:
In 2009, they conducted an online survey with 869 software developers to investigate ad-hoc reuse of ``internet code''~\citep{SojerHenkel2011}.
Even at that time, about one year after SO's launch, reuse of such code from internet sources was an essential part of developers' work. 
Our study has shown that it is common for developers to copy and paste code from SO into their projects without providing the required attribution.
Moreover, we found that developers are not aware of SO's license.
An interesting direction for future work would be to analyze if developers' usage of code snippets from the web, particularly from SO, decreases with an increase of awareness and knowledge about when code is copyrightable and which implications certain licenses have. 

Regarding code snippets on SO, Yang et al.~\citep{YangHussainOthers2016} found that Python and JavaScript snippets are more usable in terms of parsability, compilability and runnability, compared to Java and C\#.
Yang et al.~\citep{YangMartinsOthers2017} analyzed code clones between Python snippets from SO and Python projects on GH using a token-based clone detector and found a considerable number of non-trivial clones.
Abdalkareem et al. found that reusing code from SO may have a negative impact on code quality~\citep{AbdalkareemShihabOthers2017}.
%\todo{many snippets do not complile -> AST-based and other CCDs cannot be used}
Other studies aimed at identifying API usage in SO code snippets~\citep{SubramanianHolmes2013}, describing characteristics of effective code examples~\citep{NasehiSillitoOthers2012}, investigating whether SO code snippets are self-explanatory~\citep{TreudeRobillard2017}, or analyzing the impact of copied SO code snippet on application security~\citep{AcarBackesOthers2016, FischerBottingerOthers2017}.
%Brandt et al. observed the web usage of 20 students while solving a programming task and found that they spent 19\% of their time on web.
%identification of code elements~\citep{Rigby13}
%searching was most common activity for software engineers~\citep{Singer97}
Recently, Zhang et al. analyzed potential API usage violations in SO posts and found that, of the 217,818 analyzed Java and Android SO posts, 31\% may contain potential API usage violations, which could lead to program crashes or resource leaks~\citep{ZhangUpadhyayaOthers2018}.

There has also been work on the interplay between user activity on SO and GH~\citep{VasilescuFilkovOthers2013, SilvestriYangOthers2015, BadashianEstekiOthers2014}.
In particular, Vascilescu et al.~\citep{VasilescuFilkovOthers2013} showed that active GH committers ask fewer questions and provide more answers than others.
With our study, we add a new aspect to this interplay, namely how code from SO is used and attributed in GH projects.

To describe the topics of SO questions and answers, different methods like manual analysis~\citep{TreudeBarzilayOthers2011} and Latent Dirichlet Allocation (LDA)~\citep{WangLoDavidOthers2013, AllamanisSutton2013} have been used.
Automatically identifying high-quality questions and answers has been another research direction, where metrics based on the number of edits on a question~\citep{YangHauffOthers2014}, the author's popularity~\citep{PonzanelliMocciOthers2014}, and code readability~\citep{DuijnKuceraOthers2015} yielded good results.
A direction for future work is to investigate whether those high-quality questions and answers are actually referenced or used more often in GH projects.

\subsection{Licensing and Code Clones}

German and Hassan~\citep{GermanHassan2009} point to the license mismatch problem, that is combining software components with possibly conflicting licenses.
As described above, such a license conflict may arise when developers copy non-trivial code snippets from SO into their projects, because SO's license requires derivative work to use a compatible license.
An et al.~\citep{AnMloukiOthers2017} investigated whether developers respect license terms when reusing code from SO posts in a sample of 399 Android projects and found many potential license violations.
They considered a project to violate SO's license if it, among other factors, didn't ``use the CC BY-SA 3.0 or its later versions.''
However, they did not consider the compatibility of CC BY-SA 4.0 and GPL 3.0 (see Section~\ref{sec:licensing-conflicts}).
Moreover, none of the files they analyzed contained a reference ``to the corresponding Stack Overflow post.''
It is unclear if the authors also considered links to the corresponding question.
Nevertheless, these results do not contradict our estimation that in at most one quarter of the cases, code copied from SO is attributed as required (see Section~\ref{sec:conclusion}). 
% only six developers relied to their survey

A reason for such license violations may be developers struggling to understand the interaction of open source licenses.
Almeida et al. conducted an online survey with 375 software developers and found that developers struggled to understand licensing scenarios involving multiple licenses~\citep{AlmeidaMurphyOthers2018}, as it may be the case when developers want to use SO code in their projects.
As motivated above, the situation can even be more complex when code on SO is also available on other websites.
SO could address this issue by making the licensing of the content more visible on their website, and by integrating a feature that allows SO authors to easily provide an additional (more permissive) license when posting code on SO.

German et al.~\citep{GermanDiPentaOthers2009} analyzed how code siblings, i.e., code clones that evolve in a different system than the original code, flow between systems with different licenses; Gharehyazie et al.~\citep{GharehyazieRayOthers2017} and Lopes et al.~\citep{LopesMajOthers2017} found that cross-project code reuse on GH is common.
Tracing the flow of siblings between GH projects, posts on SO, and external sources is another possible direction for future work.

Two fields related to our study are source code plagiarism detection~\citep{LancasterCulwin2004} and code clone detection~\citep{RoyCordyOthers2009}, which both rely on determining the similarity of code fragments. %~\citep{Beth14}.
One of the most often cited tools for code plagiarism detection is \emph{JPlag}~\citep{PrecheltMalpohlOthers2002, BurrowsTahaghoghiOthers2007}, which uses the same algorithm to determine token string similarity like \emph{CPD}~\citep{MartinsFonteOthers2014}, the code clone detector we used in the second phase of our study.
There has been recent work on scaling the detection of code clones to large source code corpora~\citep{Ragkhitwetsagul2016, SajnaniSainiOthers2016, BurrowsTahaghoghiOthers2007} that we can build upon to be able to search for copies of all non-trivial SO code snippets in all public GH projects.

\section{Conclusion}
\label{sec:conclusion}

%\todo{private vs. work-related usage of gh and so aus beiden surveys aufgreifen}
%\todo{Was lernen wir insbesondere aus RQ3?}
%\todo{Based on the knowledge from our research, we want to build tool support to help developers maintaining copied code.}
%\todo{help SO contributors to analyze the impact of their code posted}
%\todo{Get licenses for matched repos: \url{https://developer.github.com/v3/licenses/}}

Our main goal was to quantify the amount of unattributed usages of code snippets from Stack Overflow (SO) in GitHub (GH) projects.
In a preliminary survey, half of the participants answered that they did not attribute snippets copied from SO.
However, our quantitative analysis shows that, for Java, at most a quarter of the usages of SO snippets are attributed.
We used three different approaches to find unattributed usages, always chose conservative estimates, and tried to remove as many false positive results as possible.
In the first phase, we searched for unattributed usages of the snippets from the ten most frequently referenced SO Java answers in all Java files in the \textit{BigQuery GH data set} and found that only 23\% of the copies had been attributed.
In the second phase, we utilized the token-based code clone detector CPD to find clones of a sample of 222 SO Java snippets in a sample of 2,313 popular GH Java projects and found that only 24\% of the snippet clones included a reference to SO.
In the last phase, we searched for exact copies of 29,370 SO Java snippets in 64,281 GH projects and found that only 8\% of the copies were attributed.
Thus, we think that one quarter is a reasonable upper bound for the ratio of attributed usages.
The higher ratio in the preliminary survey could be explained with a social desirability bias~\citep{Nederhof1985} affecting the respondents.

Our preliminary survey yielded that, if content from SO is attributed, developers usually add a link to the question or answer in a source code comment.
We analyzed how often these URLs are present in Java files and found that developers more often refer to questions, i.e., the whole thread, than to specific answers.
Adding a reference to a specific answer instead of the question could help to increase maintainability.
For example, one could later on check whether this answer is still the accepted one or whether a bug fix has been posted.
However, there may be cases when the question is more appropriate, e.g., when a developer wants to refer to a controversially discussed topic or a conceptual issue.
Analyzing when developers link to questions and when to answers is a direction for future work.
%We consider an analysis of the use cases for questions and answer to be future work.
%\todo{evolution of references: amount/attribution ratio/evolution of copied answer}
%\todo{code clone management auf snippets uebertragen, z.B. Benachrichtigung falls neue (akzeptierte) Antwort, oder Antwort mit hohem Score? Wie oft änder sich Antworten auf SO?}

In the three phases of our research, between 3.3\% and 11.9\% of the analyzed repositories contained a file with a reference to SO.
The popular projects from phase two were more likely to contain a reference than the broader samples of phases 1 and 3.
Depending on the project's license, the \textit{share-alike} requirement of CC BY-SA 3.0 may lead to licensing issues for those projects.
%We plan to reach out to projects having many reference to SO to investigate their awareness of the legal situation concerning the usage of code snippets from SO. % \todo{e.g. survey, open issues in projects with many copied snippets}
%\todo{survey with devs about awareness of licensing issues?}
%\todo{which repos contain many references/copies? Contact devs and point to licensing issues.}
Our second survey has shown that many developers admit copying code from SO without attribution and are not aware of the licensing and its implications.
Moreover, we found that at most 1.8\% of the GH projects with copies of SO code snippets attributed the copy and had a license that would allow a CC BY-SA 3.0-compatible usage of the SO content.
The discussions on SO about a new code license show that developers care about this topic, yet many developers do not attribute code they copy from SO posts. 
A direction for future research is to investigate this dichotomy.

%The current license for code on SO requires attribution, which, according to our results, is not done in at least two thirds of the cases.

%\todo{which repos contain many references/copies? Contact devs and point to licensing issues.}
%Depending on the context of a project, this may lead to legal issues.
%\todo{automate regex generation?}
%\todo{automate extraction of snippets}
%\todo{Maybe automatic configuration depending on number of tokens in snippets? different configurations for different lengths of snippets?}
%\todo{automation of CPD config: cluster snippets according to token length and apply different lengths with different configs}
%\todo{CPD is Open source, adapt it to only find SxF matches to optimize runtime and scalability}
%\todo{In future work, we will either adapt our filter constraint to match more generic Java answers, or specifically look for projects using the libraries from the snippets.}

The next steps of our research are to automate and scale the extraction of copyable snippets form SO and the detection of unattributed usages in GH projects.
The `reverse engineering' of the missing link to SO can help developers mitigating possible \textit{maintenance} and \textit{legal issues}, as motivated in the introduction. 
Further, using SO's official data dump, we build a data set with the extracted version history of all SO code snippets~\citep{BaltesDumaniOthers2018, BaltesDumani2018f}.
We plan to use this data set to identify buggy revisions, and then search for copies of those revisions to warn developers who copied buggy code. 
We also want to expand our analysis to other programming languages and further investigate the relations between code snippets on SO, their copies on GH, and external sources.

%We want to improve the filter criteria for retrieving SO answers that are likely to be copied and then automatically extract the contained snippets.
%Depending on their token count, one could use different CPD configurations for different classes of snippets.

%\todo{analyze clones between SO snippets}

%\todo{use commits to identify age of unreferenced usages}
%\todo{Wie viele der Referenzen enthalten "non-trivial" code snippets, also snippets, die man kopieren koennte?}
%\todo{use other ccds, e.g. AST-based (problem: snippets not nececarily syntactically correct, refer to paper from related work)}
%\todo{"impact" von snippets auf SO}
%\todo{lisencse:  "indicate if changes were made"}

%\todo{http://codereview.stackexchange.com/}
%\todo{http://codegolf.stackexchange.com/}

%todo{Kopien innerhalb von SO und von externen Quellen in SO (insbesondere in Fragen) ist Future Work.}

\begin{acknowledgements}
The authors would like to thank the participants of the online surveys, the anonymous reviewers, and Bernhard Baltes-G{\"o}tz for their valuable feedback.
Moreover, we thank Richard Kiefer for his help with the calibration of CPD and the extraction of the snippet sets and Florian Reitz for his help with database-related issues.
\end{acknowledgements}

%Your text comes here. Separate text sections with
%\section{Section title}
%\label{sec:1}
%Text with citations \citep{RefB} and \citep{RefJ}.
%\subsection{Subsection title}
%\label{sec:2}
%as required. Don't forget to give each section
%and subsection a unique label (see Sect.~\ref{sec:1}).
%\paragraph{Paragraph headings} Use paragraph headings as needed.
%\begin{equation}
%a^2+b^2=c^2
%\end{equation}

% BibTeX users please use one of
\bibliographystyle{spbasic}      % basic style, author-year citations
\bibliography{literature}   % name your BibTeX data base

\end{document}